\documentclass[12pt]{article}
\usepackage[normalem]{ulem}
\usepackage[dvipdfmx]{graphicx}
\usepackage[top=3.5cm,bottom=3.5cm,left=2cm,right=2cm]{geometry}
\usepackage{amsmath,amssymb}
\usepackage{color}
\usepackage{slashed}
\usepackage[bookmarks=true,bookmarksnumbered=true,setpagesize=false]{hyperref}

\usepackage{comment}

\newcommand{\nn}{\nonumber\\}

\newcommand{\be}{\begin{equation}}
\newcommand{\e}{\end{equation}}
\newcommand{\aln}[1]{\begin{align}#1\end{align}}

\begin{document}
\title{
\vspace{-2cm}
\vbox{
\baselineskip 14pt
\hfill \hbox{\normalsize 
}} 
\vskip 1cm
\bf \Large Cosmology of Supercooled Universe
\vskip 0.5cm
}
\author{
Kiyoharu Kawana\thanks{E-mail: \tt kawana@snu.ac.kr}
\bigskip\\
\normalsize 
\it  
$^*$  Center for Theoretical Physics, Department of Physics and Astronomy,\\
 \normalsize
\it  Seoul National University, Seoul 08826, Korea
\smallskip
}

\date{\today}

\maketitle

\begin{abstract}
First-order phase transitions (FOPTs) are ubiquitous in physics beyond the Standard Model (SM). 
Recently, models with no dimensionful parameters in the tree-level action have been attracting much attention because they can predict a very strong FOPT with ultra-supercooling. 
In this paper, we study the cosmological signatures of such a supercooling model. 
As a concrete model, we consider the SM with two additional real scalars $\phi$ and $S$, which can realize the electroweak symmetry breaking via Coleman-Weinberg mechanism.  
One of the additional scalars $S$ can naturally become a Dark Matter (DM) candidate due to the $Z_2^{}$ symmetry of the action.  
We study the FOPT of this model and calculate the Gravitational Wave (GW) signals and the thermal relic abundance of $S$ taking the filtered effects into account.   
Within the envelope approximation, we find that 
the GW peak amplitude can reach $\sim 10^{-10}$ around the frequency $f\sim 10^{-3}~$Hz for model parameters $(v_\phi^{},\lambda_{\phi S}^{})\sim (200~{\rm TeV},1.6)$ where $v_\phi^{}$ is the vacuum expectation value of $\phi$ and $\lambda_{\phi S}^{}$ is the scalar mixing coupling. 
On the other hand, the filtered DM mechanism only works for  $0.8\lesssim \lambda_{\phi S}^{}\lesssim 1$, where the GW peak amplitude is found to be quite small $\lesssim 10^{-17}$.   
%
%

\end{abstract}
\newpage

\section{Introduction}
 Observational cosmology has rapidly progressed in the last two decades, and it is now ossible to explore new physics by various observations of the Universe.  
For example, the precise measurements of the cosmic microwave background (CMB)~\cite{Planck:2018nkj,Planck:2018vyg} have revealed the magnitude of density fluctuations on large scales and the thermal history of the Universe after the recombination epoch.   
These observations place strong constraints on various inflation models~\cite{Planck:2018jri}, and it is noteworthy that a few simple inflation models with polynomial potentials are already ruled out by the CMB observations, which indicates the necessity to consider more nontrivial inflaton potentials.  
Here, we should mention that inflation models in which the Standard Model (SM) Higgs plays a role of an inflaton are  still viable and have been attracting much attention due to their phenomenological richness \cite{Bezrukov:2007ep,Bezrukov:2008ej,Bezrukov:2008ut,GarciaBellido:2008ab,Bezrukov:2010jz,Bezrukov:2012sa,Hamada:2014iga,Hamada:2014wna}.

Gravitational Wave (GW) astronomy~\cite{LIGOScientific:2016aoc,LIGOScientific:2017vwq,LIGOScientific:2018mvr} has recently become one of the most popular fields because of the possibilities of probing very early Universe physics beyond the CMB era. 
%
%
GW observations of compact binaries~\cite{LIGOScientific:2021sio,LIGOScientific:2021jlr,LIGOScientific:2017adf} and the stochastic background~\cite{NANOGrav:2020bcs,Blasi:2020mfx,Vaskonen:2020lbd,DeLuca:2020agl,Nakai:2020oit,Addazi:2020zcj,Vagnozzi:2020gtf,Benetti:2021uea} can test new physics models and gravity theories.   
In particular, the stochastic GWs from first-order phase transitions (FOPTs) have been actively studied in recent years because many new physics models predict them at the early Universe.    
%
Unfortunately, the electroweak (EW) phase transition in the pure SM is crossover~\cite{Kajantie:1995kf,Rummukainen:1998as,Csikor:1998eu}, although this motivates the study of FOPTs and GWs in the context of new physics models.    

Among various extended models, models with no dimensionful couplings in the tree-level action  
have been attracting much attention recently~\cite{Bardeen:1995kv,Meissner:2006zh,Meissner:2008gj,Foot:2007iy,Iso:2009ss,Iso:2009nw,Hur:2011sv,Iso:2012jn,Steele:2012av,Steele:2013fka,Englert:2013gz,Hashimoto:2013hta,Holthausen:2013ota,Hashimoto:2014ela,Kubo:2014ova,Endo:2015ifa,Kubo:2015cna,Brdar:2018num,Jung:2019dog} because they can naturally realize the Coleman-Weinberg (CW) mechanism and predict a very strong FOPT with ultra-supercooling~\cite{Iso:2017uuu,Hambye:2018qjv,Jinno:2016knw,Brdar:2019qut}.  
In this case, the resultant GW signals are largely enhanced, and they can be detected by future 
detectors such as LISA~\cite{LISA:2017pwj,Caprini:2015zlo,Caprini:2019egz}, 
 BBO~\cite{Corbin:2005ny}, DECIGO~\cite{Seto:2001qf,Kawamura:2006up}, Ultimate-DECIGO~\cite{Kudoh:2005as,Kawamura:2011zz}, SKA~\cite{Carilli:2004nx,Janssen:2014dka,Weltman:2018zrl} and so on. 

In this paper, we study cosmological signatures of such a classical conformal model.  
As a concrete model, we consider the most economical model studied in Refs.~\cite{Haruna:2019zeu,Hamada:2020wjh}, which can simultaneously explain the EW scale and Dark Matter (DM).\footnote{
In Ref.~\cite{Hamada:2021jls}, we have also studied the critical Higgs inflation by introducing heavy right-handed neutrinos.  
}
This model has two additional real singlet scalars $\phi$ and $S$, which is a minimum setup to realize the CW mechanism as well as the classical conformal $B$-$L$ model~\cite{Iso:2009ss,Iso:2009nw,Iso:2012jn,Iso:2017uuu}. 
From the assumption of the classical conformality, this model can realize the FOPT with ultra-supercooling and predict strong GW signals whose peak amplitude can become as large as $\sim 10^{-10}$ 
at around the frequency $\sim 10^{-3}~$Hz for some model parameters. 
For simplicity, we focus on the parameter region where only $\phi$ (and the SM Higgs) acquires the vacuum expectation value (vev) by the CW mechanism. 
In this case, $S$ can naturally become a DM candidate via usual freeze-out mechanism because its decay to SM particles is protected by the existence of $Z_2^{}$ symmetry $S\rightarrow -S$. 
However, the existence of a strong FOPT changes this picture dramatically~\cite{Baker:2019ndr,Chway:2019kft}.  
Because of the huge mass gap between the false and true vacua, only a few particles with sufficient kinetic energies can penetrate the bubble walls and survive the FOPT. 
Correspondingly, the predictions of the thermal relic abundance of $S$ are modified significantly, and we find that only the small parameter region $0.8\lesssim \lambda_{\phi S}^{}\lesssim 1$ is allowed for the coexistence between FOPT and DM abundance. 
Here, $\lambda_{\phi S}^{}$ denotes the mixing coupling between $\phi$ and $S$. 
When $\lambda_{\phi S}^{}\gtrsim 1$, the relic abundance of $S$ is typically overproduced even though a strong FOPT with ultra-supercooling is realized.  
On the other hand, when $\lambda_{\phi S}^{}\lesssim 0.8$, it becomes difficult to complete the FOPT because the bubble nucleation rate is too small.    
The resultant GW peak amplitude is predicted to be very   small $\lesssim 10^{-17}$ for such a coexisting region $0.8\lesssim \lambda_{\phi S}^{}\lesssim 1$. 
We summarize these results in Fig.~\ref{fig:GW} and the lower panel of  Fig.~\ref{fig:abundance}. 

\

The organization of this paper is as follows. 
In Section~\ref{sec:model}, we briefly review the model introduced in Refs.~\cite{Haruna:2019zeu,Hamada:2020wjh}. 
In Section~\ref{sec:FOPT}, we study the FOPT of the model.
In particular, our main focus is the calculations of the nucleation and percolation temperatures.  
In Section~\ref{sec:GW}, we calculate the GW signals based on the fitting results in Ref.~\cite{Caprini:2015zlo}.  
In Section~\ref{sec:DM}, we discuss the impacts of the strong FOPT on the relic abundance of $S$. 
Conclusions are given in Section~ \ref{sec:conclusion}.

\section{Model}\label{sec:model}
In this section, we briefly review the model introduced in Refs.~\cite{Haruna:2019zeu,Hamada:2020wjh,Hamada:2021jls}. 
We start from the following classically conformal action:
\aln{
{\cal L}={\cal L}_{\rm SM} &-\frac{1}{2}(\partial_\mu \phi)^2-\frac{1}{2}(\partial_\mu S)^2
-\lambda_H (H^\dagger H)^2
-\frac{\lambda_\phi }{4!}\phi^4-\frac{\lambda_{\phi S} }{4}\phi^2S^2-\frac{\lambda_S }{4!}S^4+\frac{\lambda_{\phi H}}{2}\phi^2(H^\dagger H)
\nn
&-\frac{\lambda_{SH} }{2}S^2(H^\dagger H)~,
\label{model}
}
where $\phi$ and $S$ are real scalars and ${\cal L}_{\rm SM}$ is the SM Lagrangian without the Higgs potential. 
%
As we will see below, the mixing coupling $\lambda_{\phi H}^{}$ is found to be very small for the successful realization of the electroweak symmetry breaking (EWSB), which justifies the neglect of the one-loop  contribution of $H$ to the effective potential of $\phi$. 
As usual, we can choose the renormalization scale $\mu$ at $\mu=M_{\rm CW}^{}$ where $\lambda_\phi^{}$ vanishes.  
As a result, all the coupling constants in Eq.~(\ref{model}) are evaluated at $\mu=M_{\rm CW}^{}$ in the following discussion.  
The one-loop effective potential of $\phi$ at zero temperature in the $\overline{\rm MS}$ scheme is 
\aln{
V_\phi^{0}(\phi)=\frac{m_S^4(\phi)}{64\pi^2}\log\left(\frac{m_S^2(\phi)}{M_{\rm CW}^2e^{3/2}}\right)+\epsilon
~,\quad m_S^2(\phi)=\frac{\lambda_{\phi S}^{}}{2}\phi^2~,
}
where $\epsilon$ is a vacuum energy constant, which has to be appropriately chosen in order to forbid the inflationary expansion at the present Universe.     
The above potential has a minimum at $m_S^2(v_\phi^{})=M_{\rm CW}^2e$, and we can eliminate $M_{\rm CW}^{}$ by using this relation as
\aln{
V_\phi^{0}(\phi)=\frac{m_S^4(\phi)}{64\pi^2}\log\left(\frac{m_S^2(\phi)}{m_S^2(v_\phi^{})e^{1/2}}\right)+\epsilon
=\frac{m_S^4(\phi)}{64\pi^2}\log\left(\frac{\phi^2}{v_\phi^2e^{1/2}}\right)+\epsilon
~,\label{CW potential}
}
which is the well known form of the CW potential~\cite{Coleman:1973jx,Kawai:2021lam}. 
The mass of $\phi$ at the true vacuum is
\aln{m_\phi^2=\frac{d^2V_\phi^0}{d\phi^2}\bigg|_{\phi=v_\phi^{}}^{}=\frac{\lambda_{\phi S}^{}}{16\pi^2}m_S^2(v_\phi^{})~.\label{phi mass}
}
The vacuum energy constant $\epsilon$ is fixed to be\footnote{To be precise, all the fields in the model contribute to the vacuum energy, but the most dominant one is the contribution by $S$ as long as $v_\phi^{}\gtrsim 1~$TeV and $\lambda_{\phi S}^{}\sim 1$.    
Note that the effect of the vacuum energy on the expansion of the Universe is negligible until $T\sim  T_\varepsilon^{}$ because the radiation energy is dominant.  
For simplicity, we also assume that the inflaton energy density is much larger than $\epsilon$. 
}
\aln{\epsilon=\frac{m_S^4(v_\phi^{})}{128\pi^2}~.\label{VTI}
}
%
In the following, we represents the Hubble scale determined by $\epsilon$ as
\aln{H_{\epsilon}^{}=\left(\frac{\epsilon}{3M_{\rm Pl}^2}\right)^{1/2}~.\label{HTI}
}
The corresponding temperature at which the radiation energy becomes equal to $\epsilon$ is given by
\aln{
T_{\epsilon}=\left(\frac{30 \epsilon}{\pi^2 g_{\epsilon}^{}}\right)^{1/4}=\frac{15^{1/4}}{2^{3/2}g_\epsilon^{1/4}}\frac{m_S^{}(v_\phi^{})}{\pi}=0.07\times \left(\frac{100}{g_\epsilon}\right)^{1/4}m_S^{}(v_\phi^{})~,\label{Tepsilon}
}
where $g_\epsilon^{}$ is the effective number of degrees of freedom (dof) at this moment. 
 
%
The CW mechanism of $\phi$ triggers the EW symmetry breaking via   $\phi^2(H^\dagger H)$ term as
\aln{
\frac{v}{v_\phi^{}}=\sqrt{\frac{\lambda_{\phi H}^{}}{2\lambda_H^{}}}~,\label{vev}
}
where $v\simeq 246$ GeV. 
This relation shows that $\lambda_{\phi H}^{}$ is quite small as long as  $v_\phi^{}\gtrsim 1~$TeV. 

In this model, the thermal relic abundance of $S$ can account for the whole DM abundance when the following relation is satisfied~\cite{Hamada:2020wjh}:
\aln{
&4\lambda_{SH}^2+\lambda_{\phi S}^2=\left(\frac{m_S^{}}{m_{\rm th} }\right)^2,\quad m_{\rm th} =1590\pm40~\text{GeV}~, 
\label{fitting}
} 
where $m_S^{}$ is the mass of $S$:
\aln{m_S^2=\frac{\lambda_{SH}^{}}{2}v^2+\frac{\lambda_{\phi S}^{}}{2}v_\phi^2~. 
}
The above relation provides a contour of $v_\phi^{}$ in the $m_S$ vs $\lambda_{SH}$ plane. 
However, the relic abundance of $S$ is drastically modified when the Universe experiences an ultra-supercooling epoch.   
We will discuss it in Section~\ref{sec:DM}.  

\section{FOPT with Ultra-Supercooling}\label{sec:FOPT}
In this section, we study the FOPT in the model introduced in the previous section. 
Thanks to the assumption of the classical conformality, the effective mass of $\phi$ around the origin is always positive at finite temperatures, which guarantees the existence of the potential barrier even at $T\ll T_c^{}$, where $T_c^{}$ is the critical temperature. 
As a result, the Universe experiences an ultra-supercooling epoch, and very strong GWs can be produced.  
In subsection~\ref{sec:finite temperature potential}, we will discuss the finite temperature effective potential of $\phi$. 
Then, we will study the nucleation and percolation in subsection~\ref{sec:Nucleation and Percolation}.

\subsection{One-loop finite temperature potential}\label{sec:finite temperature potential}
\begin{figure}
\begin{center}
\includegraphics[width=8cm]{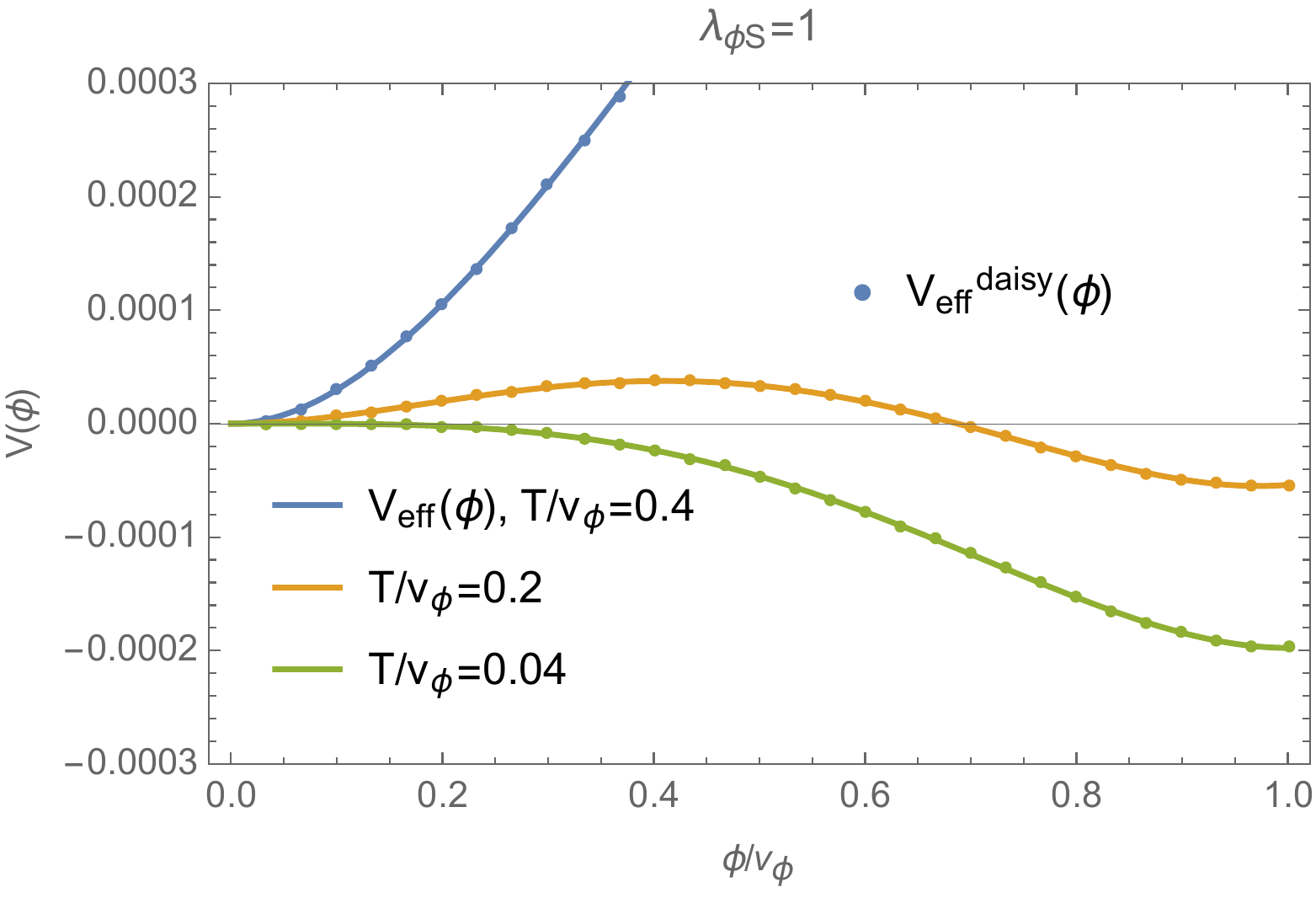}
\includegraphics[width=8cm]{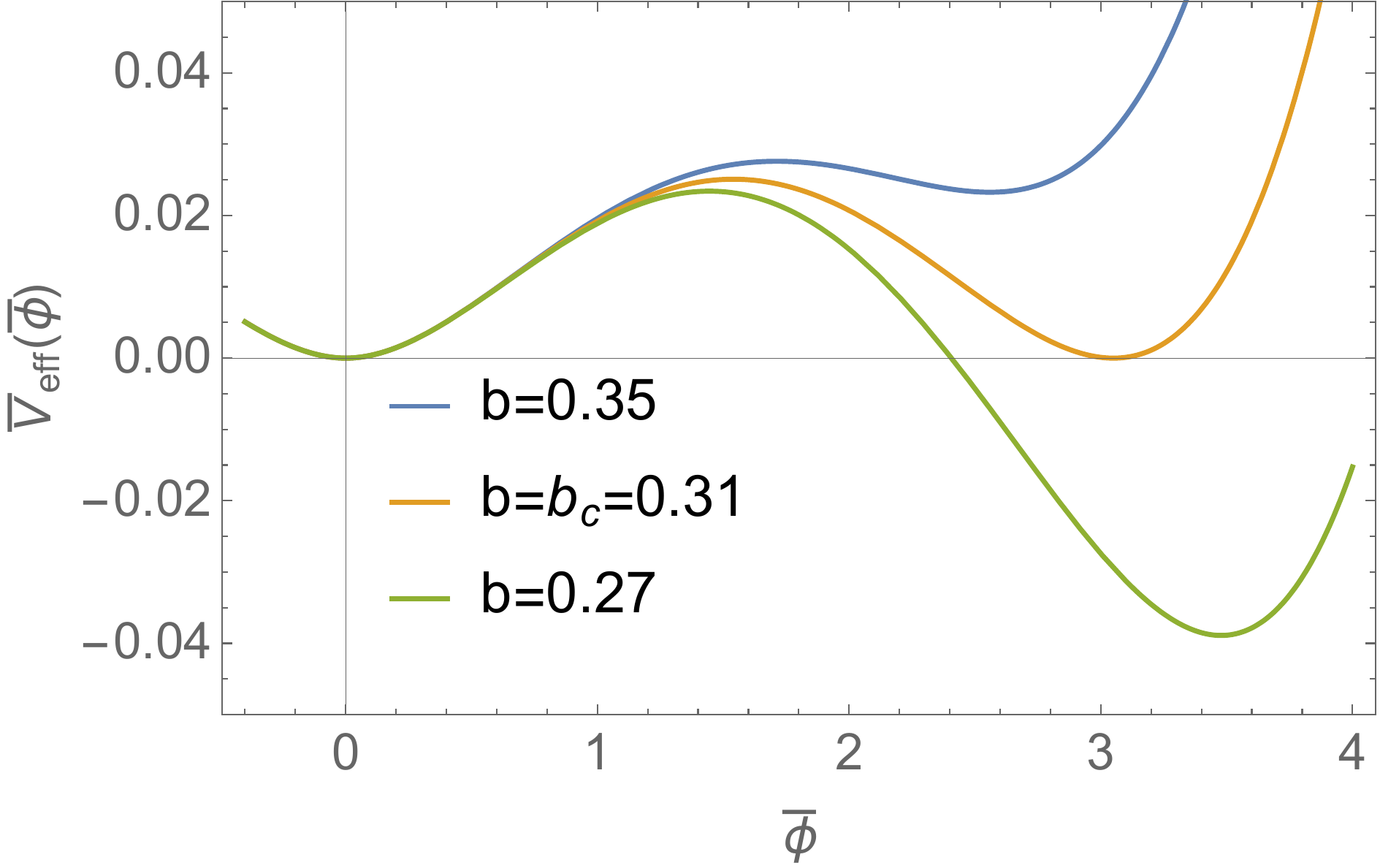}
\end{center}
\caption{Left: Comparison between $V_{\rm eff}^{}(\phi,T)$ (lines) and $V_{\rm eff}^{\rm daisy}(\phi,T)$ (points). 
Right: The normalized effective potential $\overline{V}_{\rm eff}^{}(\overline{\phi},b)$. }
\label{fig:daisy}
\end{figure}

The dynamics of FOPT is determined by the effective potential at finite temperatures. 
In general, the one-loop finite temperature corrections are 
\aln{
 V^{\rm 1loop}_{T}(\phi)
&=\sum_{i}\frac{g_i^{}T^4}{2\pi^2}\int_0^\infty dxx^2\log\left(1-e^{-\sqrt{(m_i^{}(\phi)/T)^2+x^2}}\right)-\sum_{j}\frac{f_j^{}T^4}{2\pi^2}\int_0^\infty dxx^2\log\left(1+e^{- \sqrt{(m_j^{}(\phi)/T)^2+x^2}}\right)
\\
&=\frac{T^4}{2\pi^2}\left[\sum_{i}g_i^{}I_B^{}((m_i^{}(\phi)/T)^2)-\sum_{j}f_j^{}I_F^{}((m_j^{}(\phi)/T)^2)\right]~,
}
where $g_i^{}~(f_i^{})$ represents the effective number of dof of boson (fermion) species $i$. 
In the present model, $H$ and $S$ contribute to the one-loop effective potential of $\phi$.  
However, the contribution by $H$ is negligible because the mixing coupling $\lambda_{\phi H}^{}$ is small for $v_\phi^{}\gtrsim 1~$TeV.  
Thus, only the contribution by $S$ is dominant:   
\aln{
V_{\rm eff}^{}(\phi,T)=V_\phi^{0}(\phi)+\frac{T^4}{2\pi^2}\int_0^\infty dxx^2\log\left(1-e^{-\sqrt{(m_S^{}(\phi)/T)^2+x^2}}\right)~.
}
At high temperatures, the perturbative calculations can  break down, and it is necessary to sum up all the relevant diagrams~\cite{Curtin:2016urg,Senaha:2020mop}. 
The so-called daisy resummation in the Parwani scheme~\cite{Parwani:1991gq} gives 
\aln{V_{\rm eff}^{\rm daisy}(\phi,T)=V_\phi^{0}(\phi)+\frac{T^4}{2\pi^2}\int_0^\infty dxx^2\log\left(1-e^{-\sqrt{(m_S^{\rm daisy}(\phi,T)/T)^2+x^2}}\right)~,
}
where
\aln{m_S^{\rm daisy}(\phi,T)^2&=m_S^{2}(\phi)+\frac{\lambda_{\phi S}^{}T^2}{2}I\left(m_S^{2}(\phi)/T^2+\frac{\lambda_{\phi S}^{}}{24}\right)~,
\\
I(\alpha)&=\frac{1}{\pi^2}\frac{\partial}{\partial \alpha}I_B^{}(\alpha)=\frac{1}{12}-\frac{\alpha^{1/2}}{4\pi}+\cdots~.
}
In the supercooling case, however, the effect of the daisy resummation is less important because the phase transition occurs well below the critical temperature. 
%
One can actually confirm this in the left panel in Fig.~\ref{fig:daisy} where $V_{\rm eff}^{}(\phi,T)$ (lines) and $V_{\rm eff}^{\rm daisy}(\phi,T)$ (points) are plotted for different values of $T$.  
The difference between $V_{\rm eff}^{}(\phi)$ and $V_{\rm eff}^{\rm daisy}(\phi)$ is less than 1$\%$ even when $T/v_\phi^{}=0.4$. 
Thus, we will simply study the FOPT based on $V_{\rm eff}^{}(\phi,T)$ in the following discussion. 

\

It is convenient to introduce the normalized field $\overline{\phi}$ and the coordinate $\rho$ as follows:
\aln{\overline{\phi}=m_S^{}(\phi)/T~,\quad r=\frac{\sqrt{2}\rho}{\lambda_{\phi S}^{1/2}T}~.
}
The ${\cal O}(3)$ symmetric (bounce) action can be written as
\aln{
S=
\frac{2^{7/2}\pi}{\lambda_{\phi S}^{3/2}}
\int_0^\infty d\rho\rho^2\left[\frac{1}{2}\left(\frac{d\overline{\phi}}{d\rho}\right)^2+\overline{V}_{\rm eff}^{}(\overline{\phi},b)-\overline{V}_{\rm eff}^{}(0,b)\right]:=\frac{2^{7/2}\pi}{\lambda_{\phi S}^{3/2}}f(b)~,\label{O(3) action}
}
where 
\aln{\overline{V}_{\rm eff}^{}(\overline{\phi},b)=\frac{\overline{\phi}^4}{64\pi^2}\log\left(b^2\frac{\overline{\phi}^2}{e^{1/2}}\right)+\frac{1}{2\pi^2}\int_0^\infty dxx^2\log\left(1-e^{-\sqrt{\overline{\phi}^2+x^2}}\right)~,\ b=\frac{T}{m_{S}^{}(v_\phi^{})}
~.
}
One can see that the parameter dependence of the action appears only through $b$ except for the overall normalization.  
In the following, the true vev of $\overline{V}_{\rm eff}^{}(\overline{\phi},b)$ is denoted by $\bar{v}_\phi^{}(b)$. 
Then, the critical value of $b$ is defined by 
\aln{
\overline{V}_{\rm eff}^{}(0,b_c^{})=\overline{V}_{\rm eff}^{}(\bar{v}_\phi^{}(b),b)~,
}
and it is numerically found to be
\aln{b_c^{}=0.31\quad \Leftrightarrow   \quad T_c^{}=0.31m_S^{}(v_\phi^{})~. \label{bc}
}
See the right panel in Fig.~\ref{fig:daisy} for examples, where we plot $\overline{V}_{\rm eff}^{}(\overline{\phi},b)$ for  different values of $b$. 
Eq.~(\ref{bc}) allows us to rewrite $b$ as $b=b_c^{}\times T/T_c^{}$, which implies that $f(b)$ is a mere function of $T/T_c^{}$.  
In addition, note that the orders of magnitude of $T_{\epsilon}^{}$ Eq.~(\ref{Tepsilon}) and $T_c^{}$ are not much different in the present model, which physically means that the Universe undergoes de-Sitter expansion soon after the critical temperature. 
%

\subsection{Nucleation and Percolation}\label{sec:Nucleation and Percolation}
The bubble nucleation rate per unit time per unit volume is give by
\aln{
\Gamma(T)\sim T^4\exp\left(-\frac{S_3^{}(T)}{T}\right)~,
}
where $S_3^{}(T)/T$ is the ${\cal O}(3)$ symmetric bounce action Eq.~(\ref{O(3) action}) determined by the following EOM: 
\aln{\frac{d^2\overline{\phi}}{d\rho^2}+\frac{2}{\rho}\frac{d\overline{\phi}}{d\rho}=\frac{\partial \overline{V}_{\rm eff}^{}}{\partial \overline{\phi}}~,\quad \overline{\phi}(\infty)=0~,\quad \frac{d \overline{\phi}}{d\rho}\bigg|_{\rho=0}=0~.
}
The probability of finding a point still in the false vacuum is given by $p(T)=e^{-I(T)}$ where 
\aln{
I(T)
=\int_{t_c^{}}^tdt' \Gamma(t') a(t')^3\times \frac{4\pi}{3}R(t,t')^3~,\quad 
R(t,t')=v_w^{}\int_{t'}^t\frac{ds}{a(s)}~,
}
and $v_w^{}$ is the bubble wall velocity. 
Calculation of $v_w^{}$ is one of the challenging issues in the field of phase transitions, and bubble walls are likely to run away i.e. $v_w^{}\rightarrow 1$ when a FOPT is extremely strong~\cite{Espinosa:2010hh,Wang:2020jrd,Azatov:2020ufh}.  
In this section, we simply choose $v_w^{}=1$ to discuss the percolation of the present model.   
As usual, the nucleation and percolation temperatures are defined by 
\aln{\Gamma(T_n^{})=H(T_n^{})^4~,\quad I(T_p^{})=0.3~,
}
respectively, where $H(T)$ is the Hubble scale at finite temperatures.\footnote{In the following calculations, we neglect the temperature dependence of the effective number of dof $g_{\rm eff}^{}$ because its effect is always subdominant compared to the bounce action.  
In particular, the variation of $g_{\rm eff}^{}$ is negligible in the parameter space where $T_p^{}>T_{\rm QCD}^{}$ because all the fields are relativistic in the false vacuum (i.e. symmetric phase).  
We have also numerically checked that the changes of  $T_n^{}$ and $T_p^{}$ are less than $1\%$ even when $g_{\rm eff}^{}=30$~\cite{Husdal:2016haj}.  
%
%
}
\begin{figure}[t]
\begin{center}
\includegraphics[width=8cm]{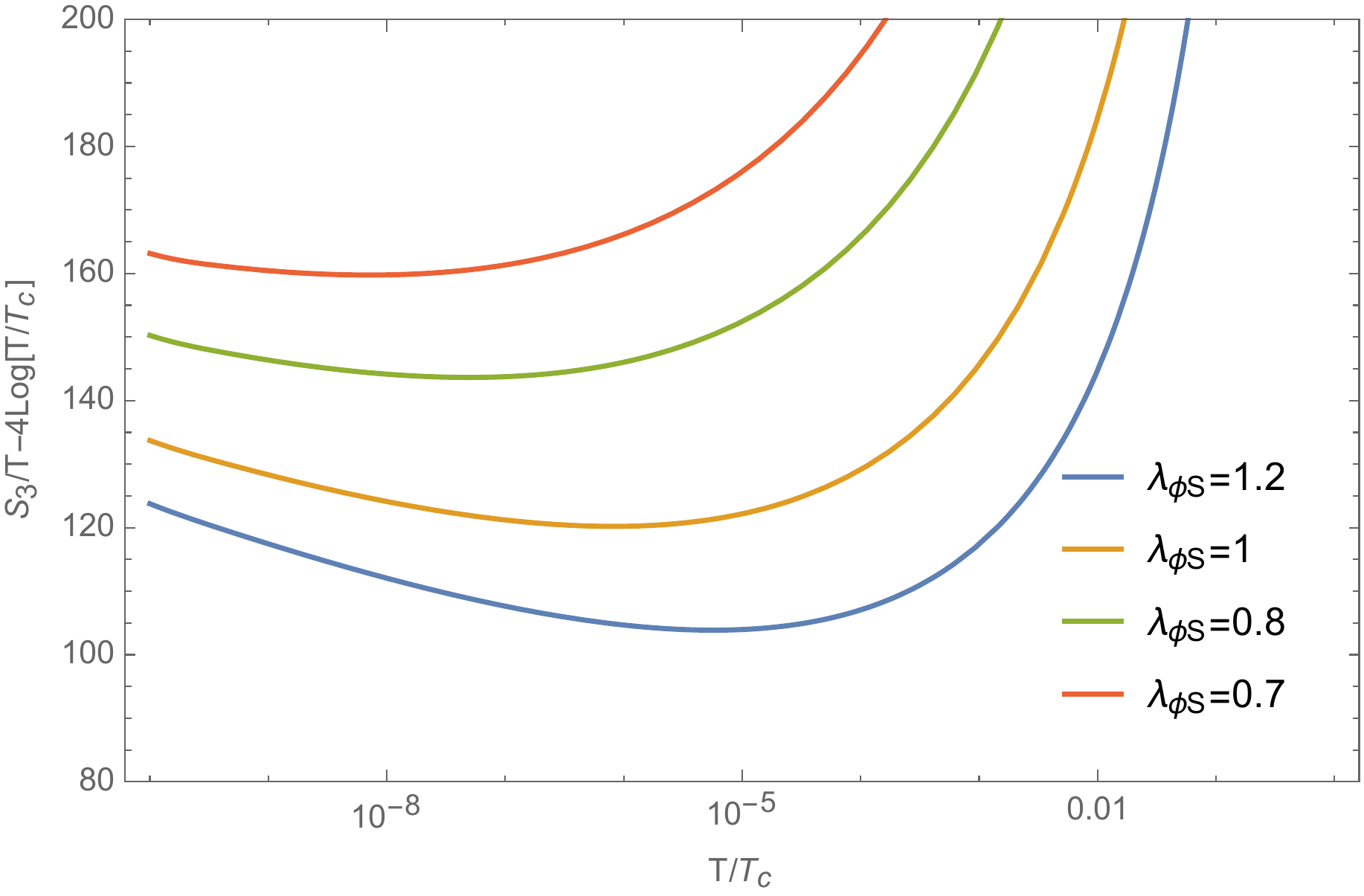}
\includegraphics[width=8cm]{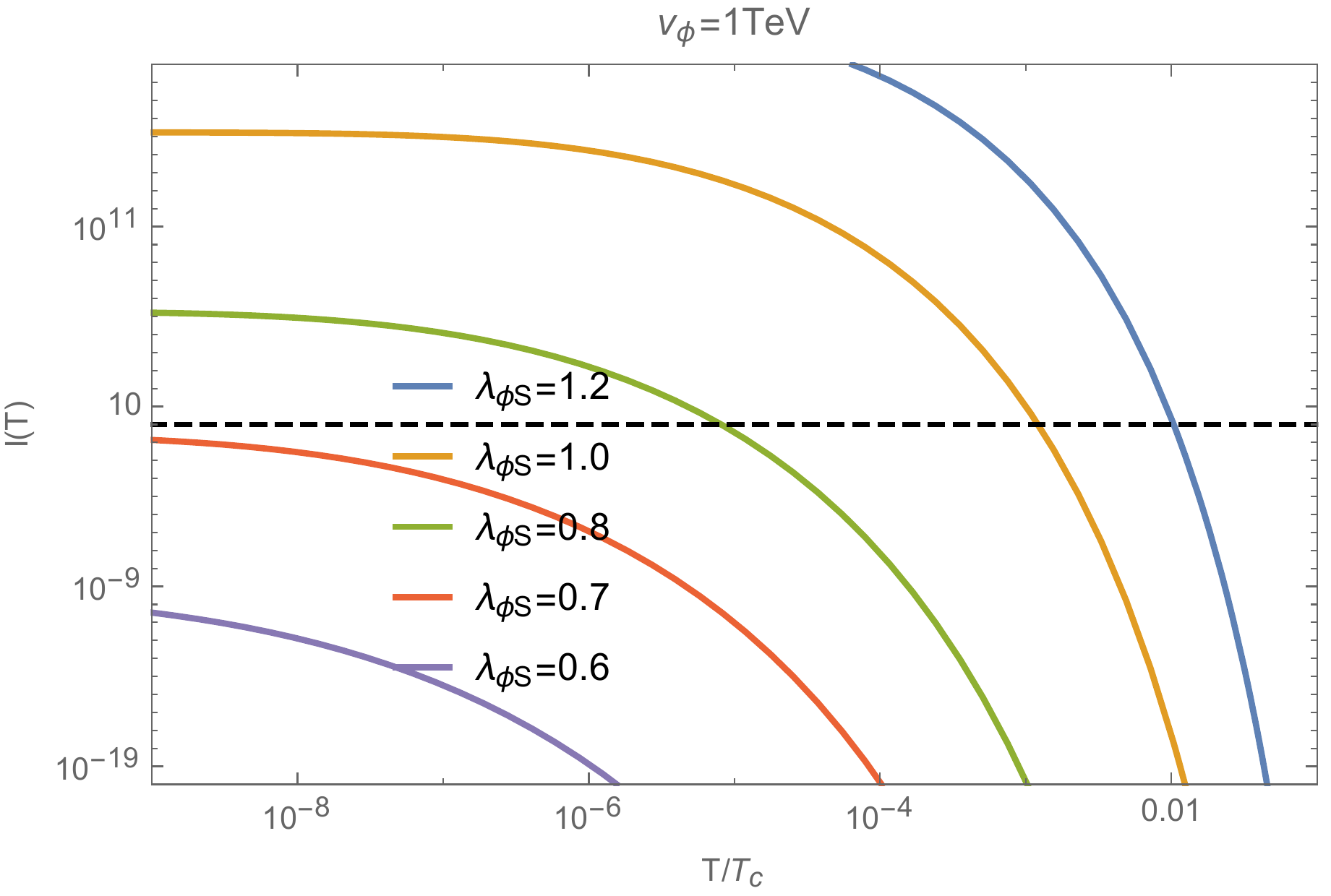}
\end{center}
\caption{Left (Right): Plots of $S_3^{}/T-4\log(T/T_c^{})~(I(T))$ as a function of $T/T_c^{}$.  
}
\label{fig:bounce}
\end{figure}

During the de-Sitter expansion $a(t)=a(t_0^{})e^{H_{\epsilon}^{}(t-t_0^{})}$, $R(t,,t')$ becomes
\aln{
R(t,t')=v_w^{}\int_{t'}^t\frac{ds}{a(s)}=-\frac{v_w^{}}{H_{\epsilon}^{}}\left(\frac{1}{a(t)}-\frac{1}{a(t')}\right)~,
}
where $H_\epsilon^{}$ is defined in Eq.~(\ref{HTI}).
By using $dT/dt=-H_{\epsilon}^{}T$, we obtain 
\aln{
I(T)
&=\frac{4\pi v_w^{3}}{3H^4_{\epsilon}}\int_{T}^{T_c^{}}\frac{dT'}{T'}\Gamma(T')\left(1-\frac{T}{T'}\right)^3
\nn
&=\frac{4\pi v_w^3}{3H^4_{\epsilon}}\int_{x}^{1}\frac{dx'}{x'}\Gamma(x')\left(1-\frac{x}{x'}\right)^3~,
}
where $x=T/T_c^{}$.  
In Fig.~\ref{fig:bounce}, we show $S_3^{}(T)/T$ (left) and $I(T)$ (right) for $v_\phi^{}=1~$TeV, where the different colors correspond to the different values of $\lambda_{\phi S}^{}$. 
One can see that $I(T)$ is quite sensitive to $\lambda_{\phi S}^{}$ because $\Gamma(T)$ exponentially depends on $S_3^{}(T)/T$. 
This sharp parameter dependence of $I(T)$ is then reflected in the sharp parameter dependence of $T_n^{}$ and $T_p^{}$.  

\

\begin{figure}
\begin{center}
\includegraphics[width=7cm]{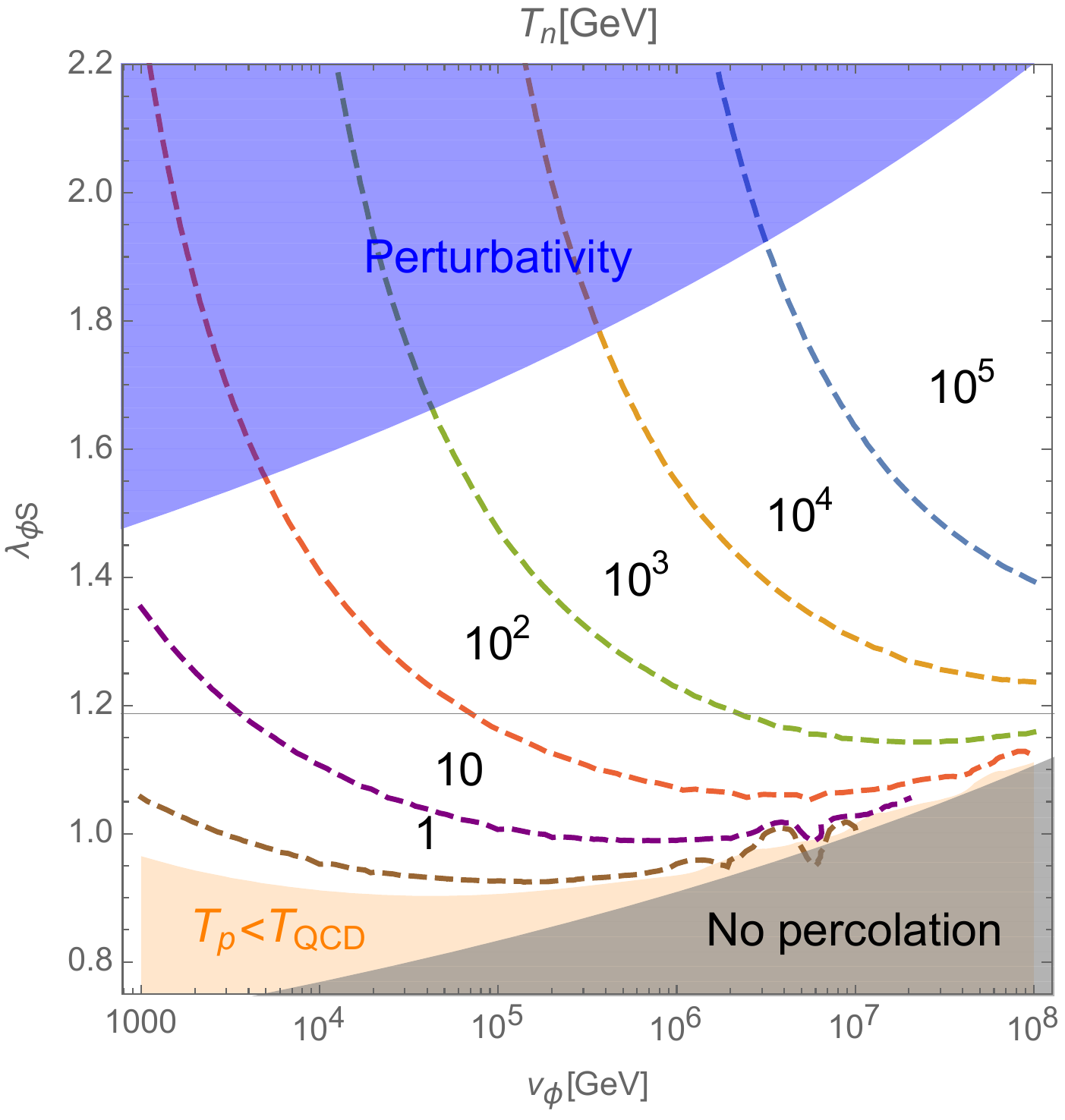}
\includegraphics[width=7cm]{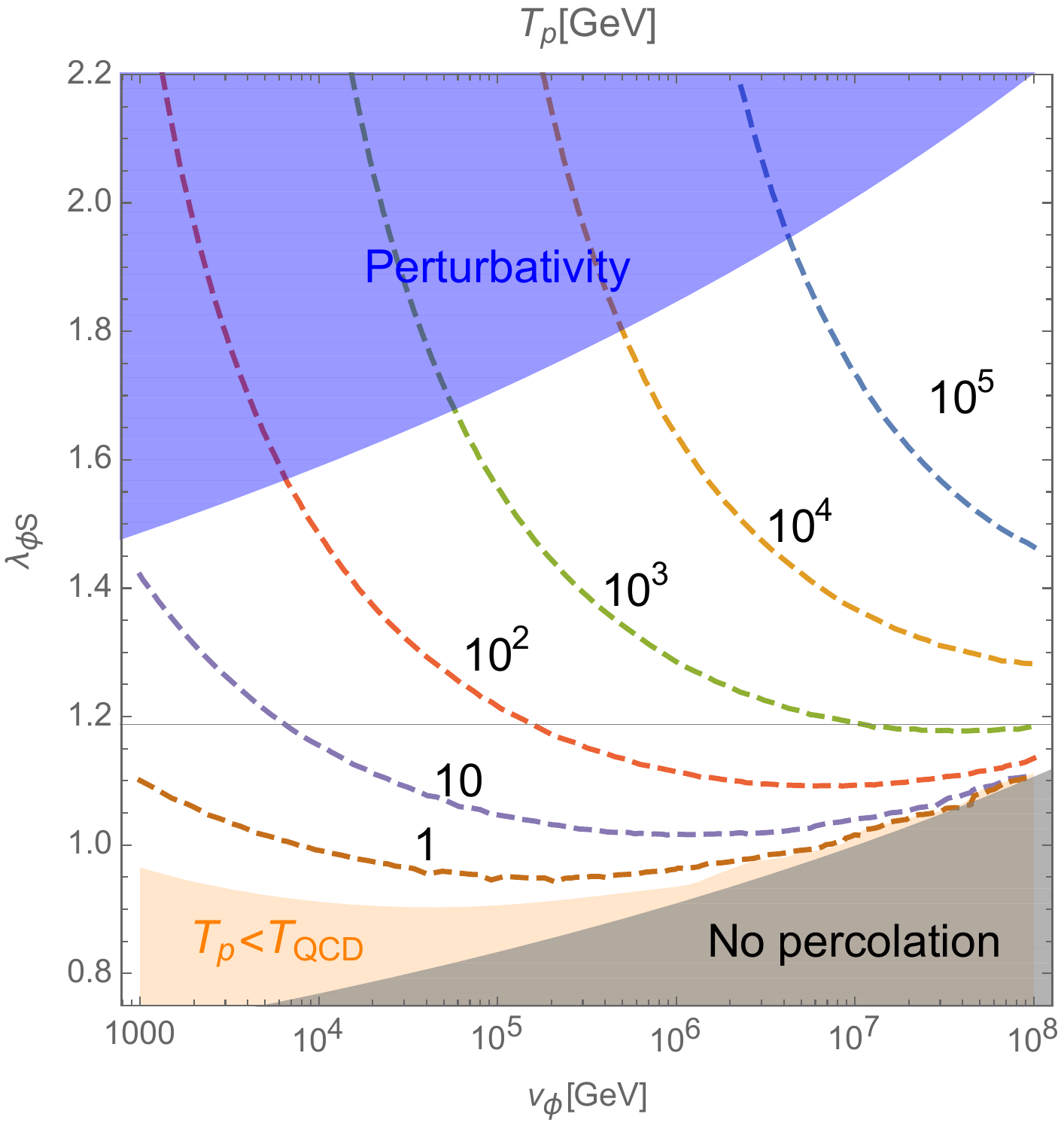}
\end{center}
\caption{Left (Right): Contours of nucleation (percolation) temperature $T_n^{}~(T_p^{})$ in the $v_\phi^{}$-$\lambda_{\phi S}^{}$ plane.
}
\label{fig:temperatures}
\end{figure}
In Fig.~\ref{fig:temperatures}, we show the contours of $T_n^{}$ (left) and $T_p^{}$ (right) in the $v_\phi^{}$-$\lambda_{\phi S}^{}$ plane. 
Here, the gray region is the parameter region where the percolation condition $I(T_p^{})=1$ can not be satisfied, and the blue region is theoretically excluded by the perturbativity (i.e. the existence of Landau pole below the Planck scale). 
The orange region corresponds to the parameter region where $T_p^{}\leq T_{\rm QCD}^{}\sim 150~$MeV, which implies that the QCD phase transition occurs during the FOPT. 
%
%
We can actually see the sharp parameter dependences     of these temperatures, and they are typically much lower than the critical temperature $T_c^{}\sim m_S^{}(v_\phi^{})\sim v_\phi^{}$. 

We should comment on the completion criterion of a FOPT with supercooling. 
Even if the percolation criterion $I(T_p^{})=1$ is satisfied, this does not necessary mean the completion of a FOPT because false vacuum regions can continue to  expand exponentially even after $T_p^{}$.   
In other words, the physical volume of the false vacuum regions $V_{\rm false}^{}\propto a^{3}(t)e^{-I(t)}$ must decrease at around $T=T_p^{}$~\cite{Turner:1992tz,Megevand:2016lpr,Wang:2020jrd}. 
This condition is equivalent to
\aln{\frac{1}{V_{\rm false}^{}}\frac{dV_{\rm false}^{}}{dt}=3H(T)-\frac{dI(t)}{dt}=H(T)\left(3+T\frac{dI(T)}{dT}\right)<0~.
}
We numerically checked that the above condition is always satisfied  in the present model at $T=T_p^{}$.  
%

\subsection{EWSB triggered by QCD phase transition}
Thermal histories of the Universe that undergoes an ultra-supercooling epoch are distinguished by whether $T_p^{}$ becomes smaller or larger than the QCD temperature $T_{\rm QCD}^{}\sim 150~{\rm MeV}$~\cite{Witten:1980ez,Iso:2017uuu,vonHarling:2017yew}. 

\

\noindent $\bullet$ {$T_p^{}\gtrsim T_{\rm QCD}^{}$}\\
In this case, the FOPT is completed before the QCD phase transition.  
In our model, most of the parameter regions belong to this case  
as one can see from Fig.~\ref{fig:temperatures}. 
One of the observational signals is the productions of strong GWs, and we will study it in the next section. 
After the FOPT, there is a reheating era by the oscillations of $\phi$, which plays a very important role to determine the GW signals and the DM abundance below.   
%

\

\

\noindent $\bullet$ {$T_p^{}\lesssim T_{\rm QCD}^{}$}\\
In this case, the QCD phase transition occurs before the completion of percolation.  
At $T=T_{\rm QCD}^{}$, chiral condensation occurs~\cite{Braun:2006jd} and the Higgs potential acquires the linear term via the top Yukawa coupling as
\aln{-\frac{y_t^{}}{\sqrt{2}}\langle \bar{t}t\rangle h\sim -\frac{y_t^{}}{\sqrt{2}}\Lambda_{\rm QCD}^3 h~. 
}
Then, the minimum of the Higgs potential is shifted to 
\aln{\langle h\rangle =v_{\rm QCD}^{}\sim  (y_t^{}\langle \bar{t}t\rangle/(\sqrt{2}\lambda_H^{}))^{1/3}\sim \Lambda_{\rm QCD}^{}~.
}
The history of the Universe after $T=T_{\rm QCD}^{}$ is controlled by the $\phi$ potential around the origin. 
By using the high temperature expansion, it is given by 
\aln{\frac{T^2m_S^2(\phi)}{24}-\frac{\lambda_{\phi H}^{}}{4}v_{\rm QCD}^2\phi^2=\frac{m_S^2(v_\phi^{})}{24 v_\phi^2}
\left(T^2-6\left(\frac{m_H^{}}{m_S^{}(v_\phi^{})}\right)^2v_{\rm QCD}^2\right)\phi^2~,
}
where $m_H^{}=\sqrt{2\lambda_H^{}}\times v=125~$GeV is the Higgs mass. 
When $m_S^{}(v_\phi^{})\gtrsim \sqrt{6}m_H^{}$, the coefficient is positive at $T=T_{\rm QCD}^{}$, which means that $\phi$ continues to be trapped in the false vacuum until $T=T_p^{}$ or 
\aln{T_{\rm end}^{}=\sqrt{6}\frac{m_H^{}}{m_S^{}(v_\phi^{})}v_{\rm QCD}^{}~,
}
at which the coefficient of quadratic term of $\phi$ becomes negative. 
During such a trapping, the Universe continues to expand exponentially, and everything is diluted. 
Such a dilution after the QCD phase transition has a lot of implications in particle cosmology. 
See \cite{Iso:2020pzv,Iso:2021tuf} for examples.  
%
On the other hand, when $m_S^{}(v_\phi^{})\lesssim \sqrt{6}m_H^{}$, the scalar fields start to roll down the potential right after $T=T_{\rm QCD}^{}$, which implies that the phase transition becomes second-oder in this case. 

\

In the following sections, we will focus on the parameter region where $T_p^{}>T_{\rm QCD}^{}$, and discuss cosmological signatures of the ultra-supercooled Universe.     

\section{Gravitational Wave Signals}\label{sec:GW}
In this section, we study the GW signals produced by the FOPT. 
Because the GW energy spectrum $\Omega_{\rm GW}^{}(f)$ is a function of various parameters of a FOPT, we first clarity their definitions and how they are calculated in subsection~\ref{Various parameters}. 
Then, we calculate the GW signals in the present model based on the fitting results in Ref.~\cite{Caprini:2015zlo}.
\footnote{
Calculating the GW energy spectrum in (ultra-)supercooling case is still a controversial problem as well as the determination of wall velocity, and more dedicated studies are necessary to  predict it more accurately.  
See also Refs.~\cite{Wang:2020jrd,Gouttenoire:2021kjv,Ellis:2019oqb} and references therein.  
}
Note that, unlike moderate FOPTs, dilution effects by the reheating become very important in ultra-supercooling case.

\subsection{Various parameters
}\label{Various parameters}
\begin{figure}
\begin{center}
\includegraphics[width=7cm]{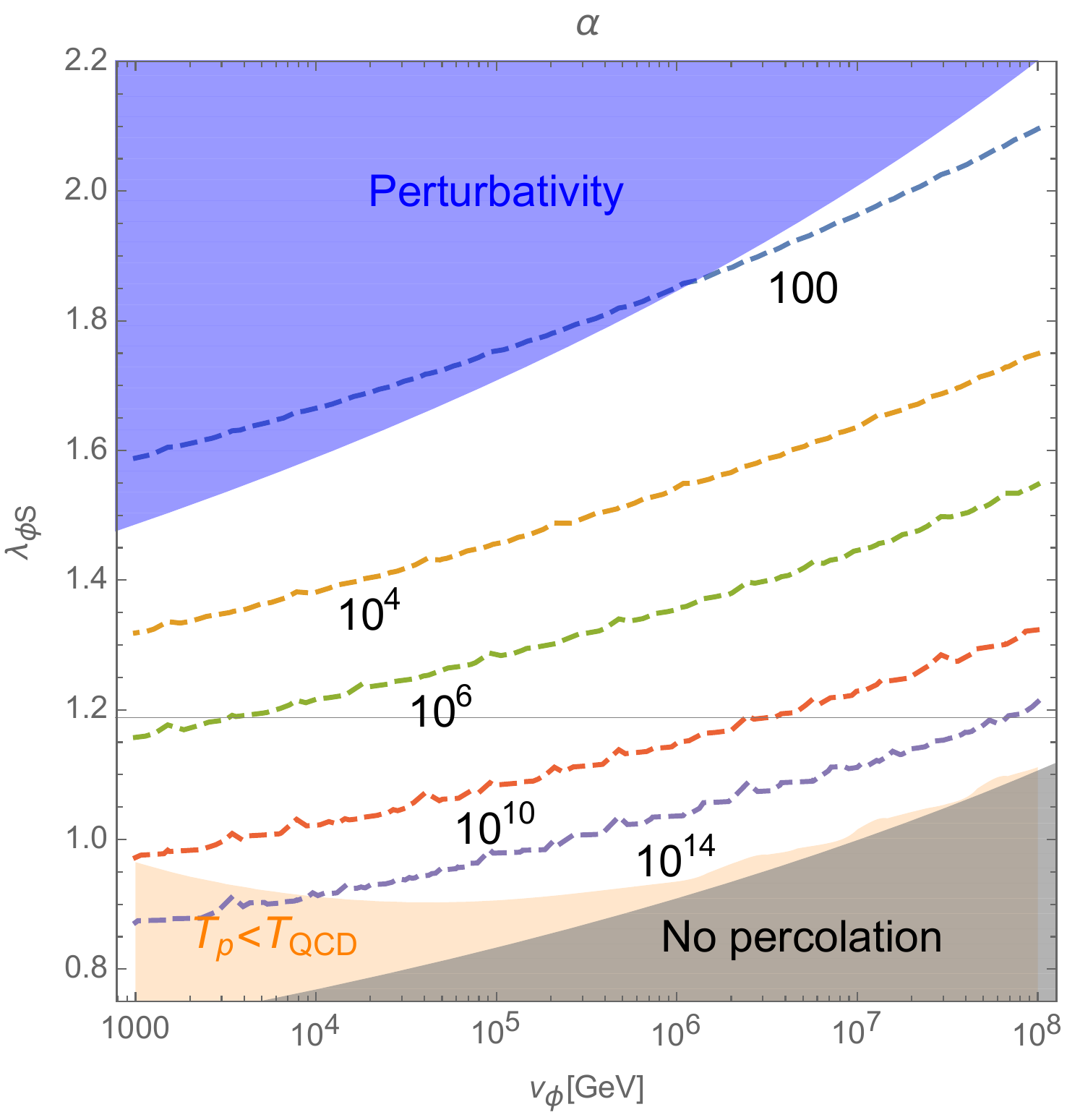}
\includegraphics[width=7cm]{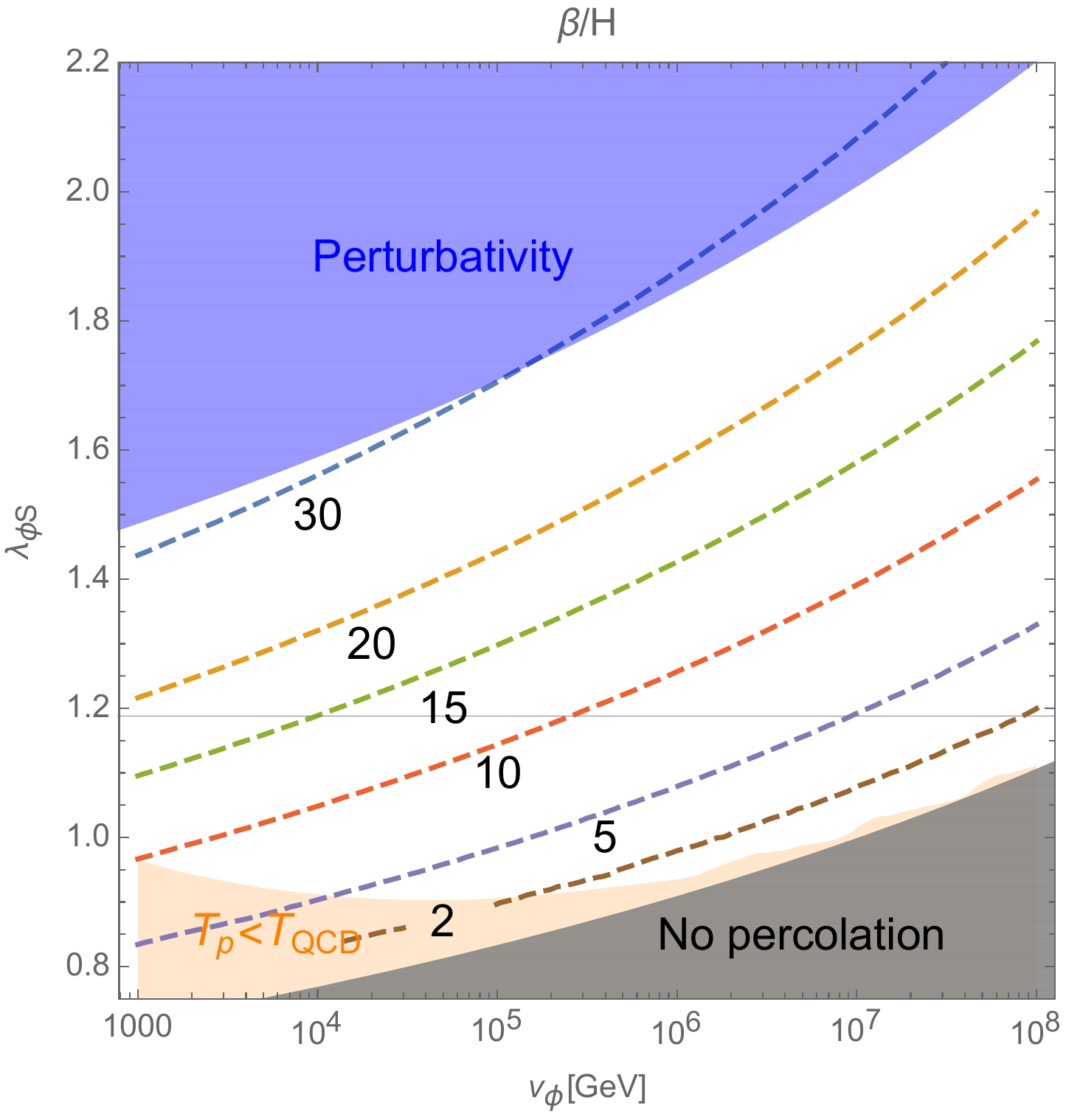}
\end{center}
\caption{Contours of $\alpha$ (left) and $\beta$ (right) in the $v_\phi^{}$-$\lambda_{\phi S}^{}$ plane. 
}
\label{fig:parameters}
\end{figure}
The GW energy spectrum $\Omega_{\rm GW}^{}(f)$ produced by a FOPT is a function of various parameters~\cite{Caprini:2015zlo,Caprini:2019egz}: phase transition strength parameters $\alpha$ and $\beta$, wall velocity $v_w^{}$, and energy efficiency factors $\kappa_\phi^{},\kappa_v^{}$.    
First, $\alpha$ is the ratio between the latent heat energy and the radiation energy:
\aln{
\alpha=\left[\Delta V_{\rm eff}^{}(T)-T\frac{\partial \Delta V_{\rm eff}^{}(T)}{\partial T}\right]\bigg/\rho_R^{}(T)~,
}
where
\aln{\Delta V_{\rm eff}^{}(T)=V_{\rm eff}^{}(\phi_{\rm false}^{},T)-V_{\rm eff}^{}(\phi_{\rm true}^{},T)~.
}
In general, the strength of a FOPT is measured by $\alpha$, and 
ultra-supercooling corresponds to $\alpha >1$~\cite{Wang:2020jrd}.  
The other strength parameter is defined by
\aln{
\beta=-H(T)T\frac{\partial \ln \Gamma (T)}{\partial T}~,
}
which determines the duration of the FOPT and the characteristic frequency of the GWs.   
In general, these parameters are functions of $T$, which gives rise to the $T$ dependence of the GW wave spectrum. 
In this paper, we evaluate $(\alpha,\beta)$ at $T=T_p^{}$,  as is commonly accepted in many literatures~\cite{Caprini:2015zlo,Jinno:2016knw,Caprini:2019egz}.  
%
Note that the nucleation temperature $T_n^{}$ is also frequently used as the temperature at which GWs are supposed to be produced, but both choices give qualitatively the same results as long as a FOPT is mild, $\alpha\ll1,~\beta/H\sim 100$.  
On the other hand, significant differences can arise in ultra-supercooling case because of the large hierarchy between  $T_n^{}$ and $T_p^{}$ ~\cite{Wang:2020jrd,Jinno:2016knw,Ellis:2019oqb}. 
%
%
In Fig.~\ref{fig:parameters}, we show the contours of $\alpha$ (left) and $\beta/H$ (right) in the $v_\phi^{}$-$\lambda_{\phi S}^{}$ plane in the present model.  
One can see that our model typically predicts ultra-supercooling $\alpha\gg1~, \beta/H={\cal O}(1\sim 30)$,  and $\alpha$ can even reach ${\cal O}(10^{14})$ for $\lambda_{\phi S}^{}\sim 1$.  
%
%

\

In principle, the wall velocity $v_w^{}$ (or corresponding Lorentz factor $\gamma=(1-v_w^2)^{-1/2}$) is determined by the balancing condition between the vacuum energy pressure and the friction pressure~\cite{Gouttenoire:2021kjv}. 
In this paper, we simply take the Jouguet detonation~\cite{Steinhardt:1981ct} 
\aln{
v_w^{}=\frac{\sqrt{\alpha(2+3\alpha)}+1}{\sqrt{3}(1+\alpha)}~ 
}
as a bench-mark and leave more detailed study for future investigation. 
As for the efficiency factors, we rely on the following runaway picture~\cite{Caprini:2015zlo}.  
As $\alpha$ is increased, the terminal velocity quickly increases and finally becomes the speed of light at some point $\alpha=\alpha_\infty^{}$, which is given by~\cite{Espinosa:2010hh}
\aln{\alpha_\infty^{}\simeq \frac{30}{24\pi^2}\frac{\sum_i c_i^{}\Delta m_i^2}{g(T)T^2}~.
} 
Here, the sum runs over all particle species $i$ that are light in the false vacuum and become heavy in the true vacuum, $\Delta m_i^2$ is the difference of their squared masses, $g(T)$ is the total effective number of dof at $T$, and $c_i^{}$ is equal to $(1/2)g_i^{}$ for boson (fermion) with $g_i^{}$ the effective number of dof.  
For $\alpha>\alpha_\infty^{}$, the fluid profile no longer changes and the surplus energy is transferred to the motion of bubble walls, and its energy fraction is represented by
\aln{\kappa_\phi^{}=1-\frac{\alpha_\infty^{}}{\alpha}~.
} 
Then, the efficiency factor $\kappa_v^{}$, which is defined by the ratio of the bulk kinetic energy over the vacuum energy, is given by~\cite{Caprini:2015zlo}
\aln{
\kappa_v^{}=\frac{\alpha_\infty^{}}{\alpha}\frac{\alpha_\infty^{}}{0.73+0.083\sqrt{\alpha_\infty^{}}+\alpha_\infty^{}}\quad (\text{runaway})~. 
\label{kappav}
}
In the present ultra-supercooling case, Eq.~(\ref{kappav}) typically becomes 
\aln{\kappa_v^{}\sim \frac{\alpha_\infty}{\alpha}\propto \frac{v_\phi^2 T^2}{\epsilon}\ll 1~, 
}
which shows that the contributions from the bulk kinetic motion are  negligible.  
As a result, the main contribution to the GW spectrum is the collisions of bubble walls in  ultra-supercooling case.   
%

\subsection{Gravitational wave spectrum}
\begin{figure}[t!]
\begin{center}
\includegraphics[width=7cm]{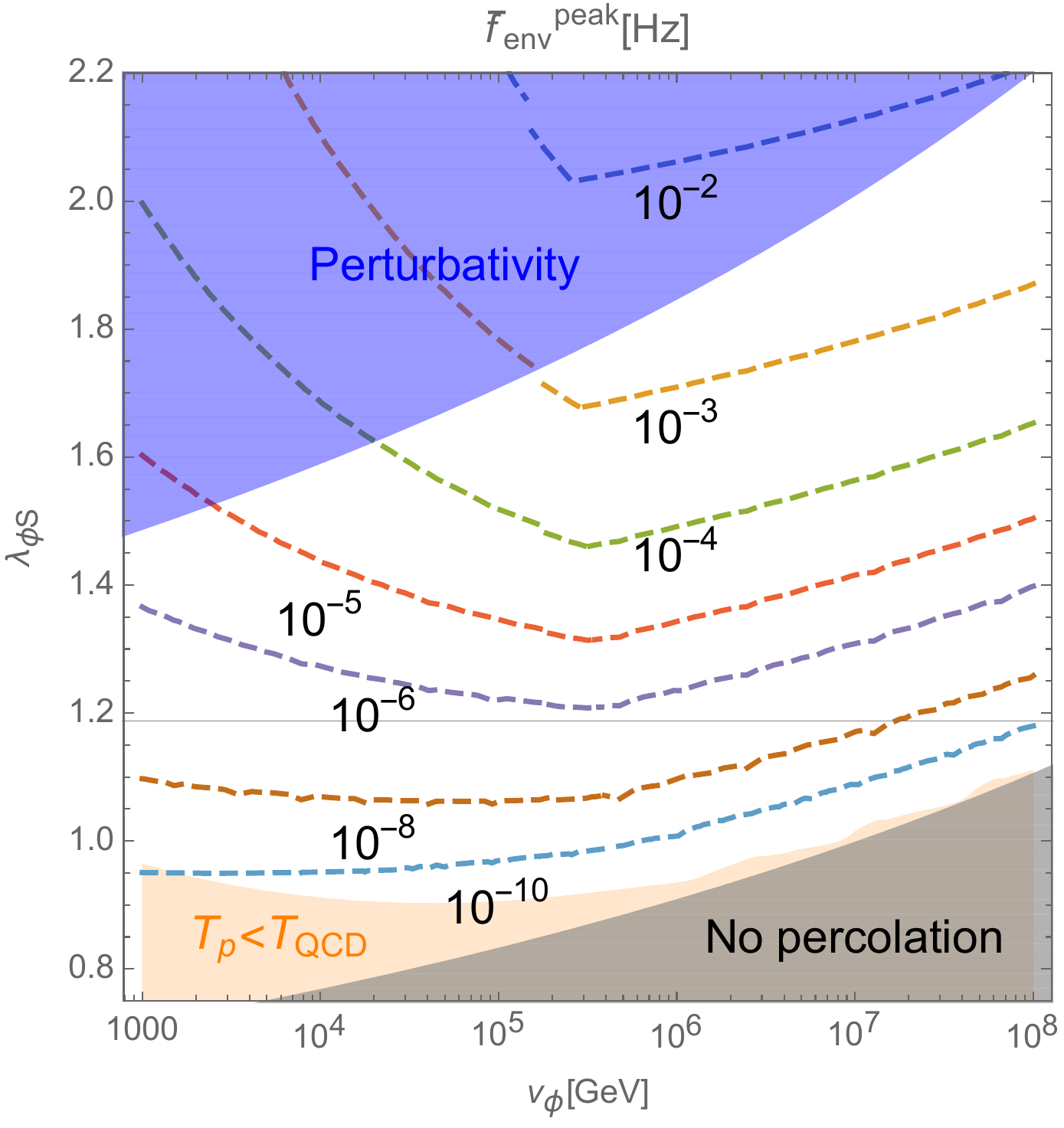}
\includegraphics[width=7cm]{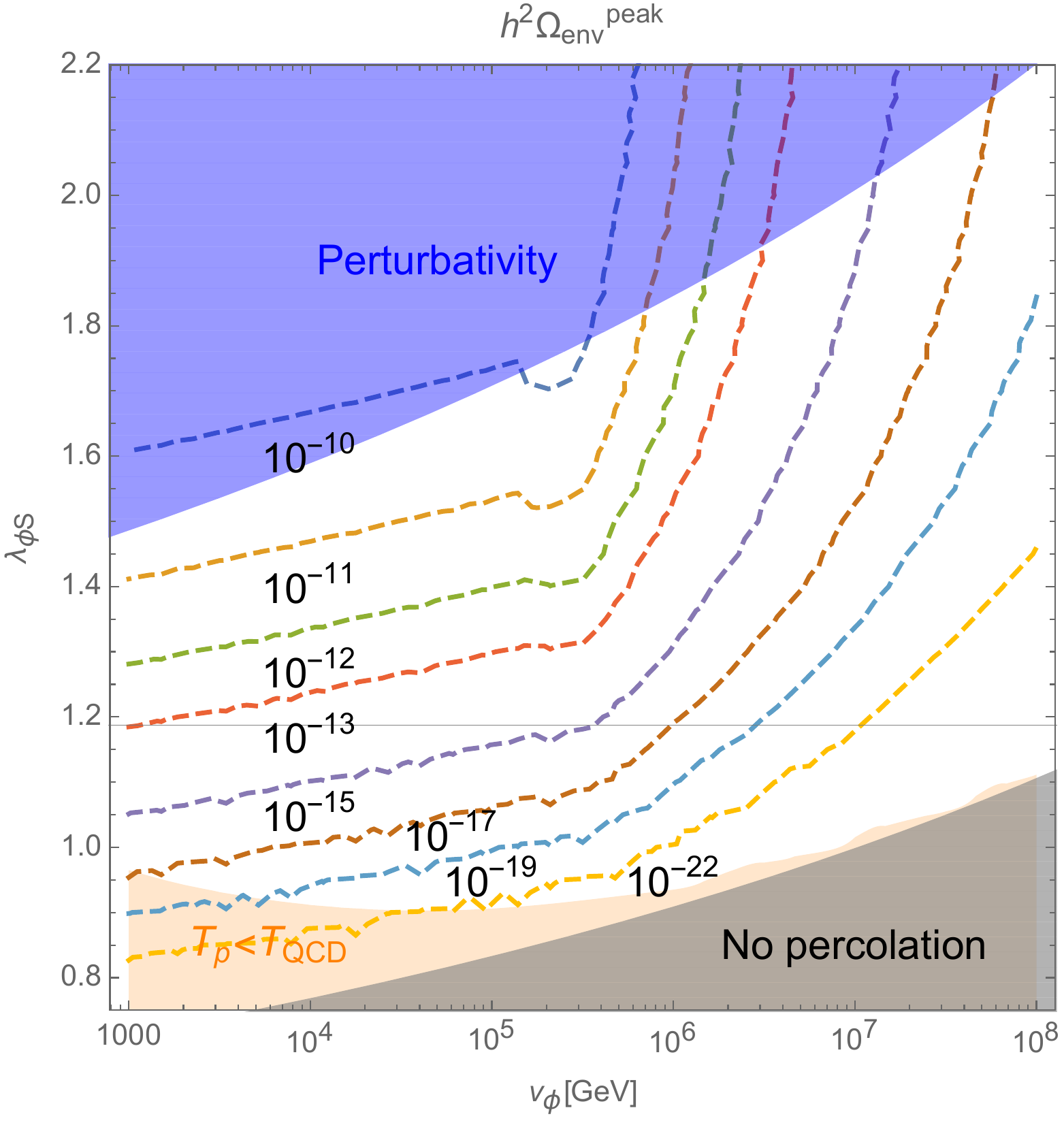}
\includegraphics[width=8.5cm]{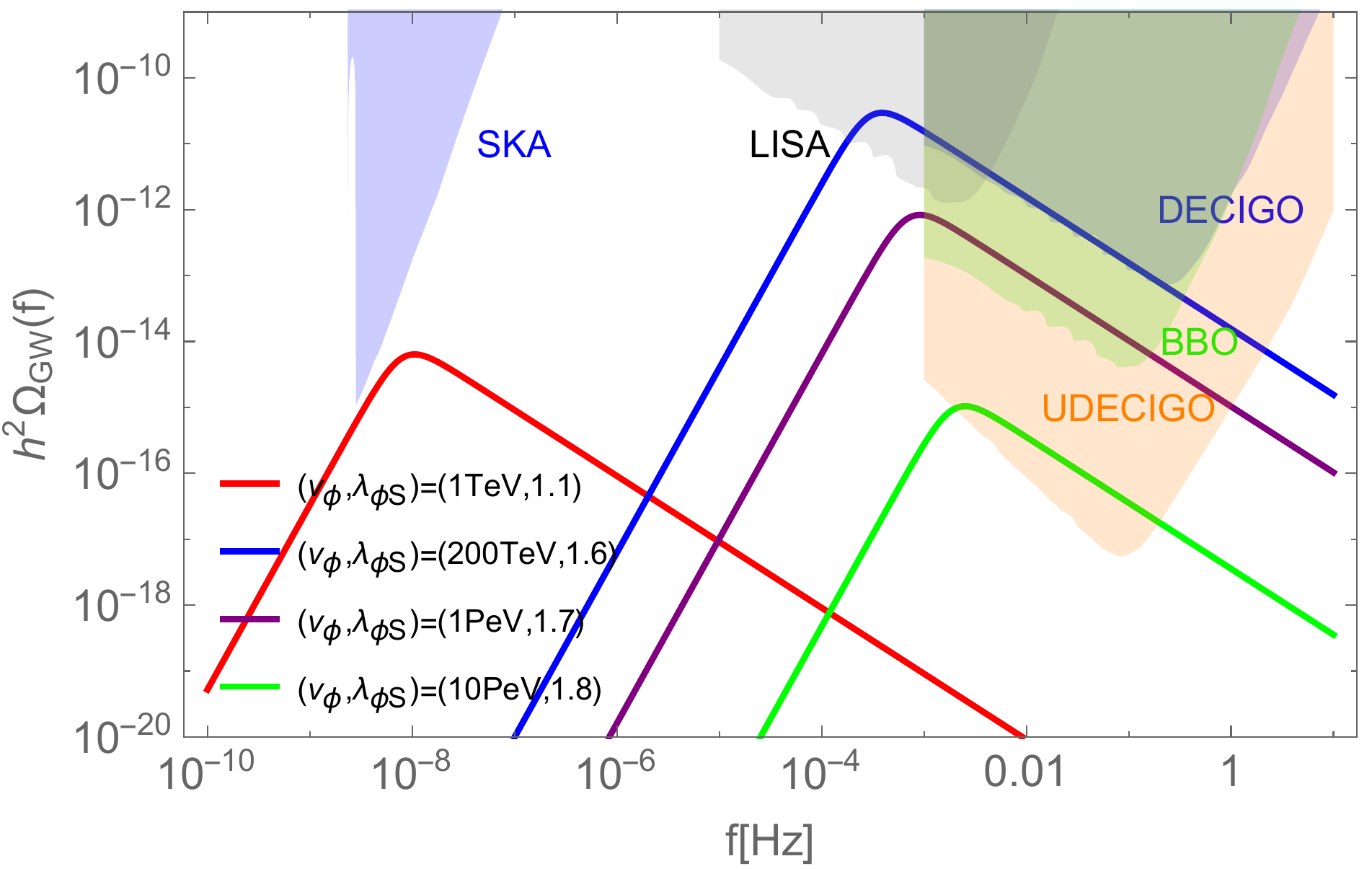}
\end{center}
\caption{Upper: Contours of $f_{\rm env}^{\rm peak}$ (left) and $h^2\Omega_{\rm env}^{\rm peak}$ (right) in the $v_\phi^{}$-$\lambda_{\phi S}^{}$ plane. 
Lower: Plots of GW energy spectra and detector sensitivities. 
}
\label{fig:GW}
\end{figure}
As we have discussed above, the main contribution to the GW spectrum is the collisions of bubble walls. 
In this paper, we use the numerical fitting results~\cite{Caprini:2015zlo} 
\aln{
h^2\Omega_{\rm env}^{}(f)=1.67\times 10^{-5}\left(\frac{H(T_p^{})}{\beta}\right)^2\left(\frac{\kappa_\phi^{}\alpha}{1+\alpha}\right)^2\left(\frac{100}{g_*^{}}\right)^{1/3}\left(\frac{0.11v_w^3}{0.42+v_w^2}\right)S_{\rm env}^{}(f)~,
}
where
\aln{S_{\rm env}^{}(f)=\frac{3.8(f/f_{\rm env}^{\rm peak})^{2.8}}{1+2.8(f/f_{\rm env}^{\rm peak})^{3.8}}~.
}
Here, $g_*^{}$ is the effective number of dof at $T=T_p^{}$. 
The peak frequency $f_{\rm env}^{\rm peak}$ is also numerically given by~\cite{Caprini:2015zlo}
\aln{f_{\rm env}^{\rm peak}=1.65\times 10^{-5}{\rm Hz}\left(\frac{0.62}{1.8-0.1v_w^{}+v_w^2}\right)\left(\frac{\beta}{H(T_p^{})}\right)\left(\frac{T_p^{}}{100~{\rm GeV}}\right)\left(\frac{g_*^{}}{100}\right)^{1/6}~.
}
Note that the above results are based on the assumption that there is no additional entropy productions after $T=T_p^{}$. 
In the present model, we have to take the reheating era after the FOPT into account. 
%
By considering such effects, the GW spectrum and the peak frequency at the present Universe become 
\aln{
h^2\overline{\Omega}_{\rm env}^{}(f)&=\rho_{c}^{-1}h^2\left(\frac{a(T_R^{})}{a(T_0^{})}\right)^4\left(\frac{a(T_p^{})}{a(T_R^{})}\right)^4\times \rho_{\rm env}(f)|_{T=T_p^{}}
\nn
&=\left(\frac{g_*^{}T_p^3}{g_R^{}T_R^3}\right)^{4/3}\left(\frac{\pi^2 g_R^{}T_R^4/30}{\epsilon}\right)^{4/3}\times h^2\Omega_{\rm env}^{}(f)|_{f_{\rm env}^{\rm peak}\rightarrow \bar{f}_{\rm env}^{\rm peak}}~,\label{OmegaGW}
\\
\bar{f}_{\rm env}^{\rm peak}&=\left(\frac{a(T_R^{})}{a(T_0^{})}\right)\left(\frac{a(T_p^{})}{a(T_R^{})}\right)\times f_{\rm env}^{\rm peak}\bigg|_{T=T_p^{}}
=\left(\frac{g_*^{}T_p^3}{g_R^{}T_R^3}\right)^{1/3}\left(\frac{\pi^2 g_R^{}T_R^4/30}{\epsilon}\right)^{1/3}f_{\rm env}^{\rm peak}~,\label{peak frequency}
}
where $T_R^{}$ is the reheating temperature and $g_{R}^{}$ is the effective number of dof at this moment. 
In general, $T_R^{}$ is determined by the decay and scattering of $\phi$. 
When such processes are fast enough, we have $\pi^2g_R^{}T_R^4/30\sim \epsilon$, and the suppression factor by the oscillation of $\phi$ disappears.  
More detailed calculations of $T_R^{}$ are presented in \ref{reheating}.

We now have all the necessary inputs to calculate the GW signals from the FOPT. 
All the parameters in Eqs.~(\ref{OmegaGW})(\ref{peak frequency}) are functions of $\lambda_{\phi S}^{}$ and $v_\phi^{}$, which allows us to plot the contours of $f_{\rm env}^{\rm peak}$ and the peak energy density $h^2\Omega_{\rm env}^{\rm peak}=h^2\overline{\Omega}_{\rm env}^{}(f_{\rm env}^{\rm peak})$ in the same way as Fig.~\ref{fig:temperatures} and \ref{fig:parameters}.  
In the upper panels in Fig.~\ref{fig:GW}, we show the contours of $f_{\rm env}^{\rm peak}$ (left) and $h^2\Omega_{\rm env}^{\rm peak}$ (right)  in the $v_\phi^{}$-$\lambda_{\phi S}^{}$ plane. 
One can see that the peak amplitude can become as large as ${\cal O}(10^{-10})$ around $(v_\phi^{},\lambda_{\phi S}^{})=(200~{\rm TeV},1.6)$ with the frequency $\sim 10^{-3}~$Hz. 
Note that even though the FOPT becomes stronger with decreasing $\lambda_{\phi S}^{}$, the GW signals are getting weaker due to the suppressions by reheating.  
This is also one of the interesting consequences of  ultra-supercooling. 

In the lower panel in Fig.~\ref{fig:GW}, we show the GW energy spectra for a few different values of model parameters where the sensitivity curves of future detectors are also shown in different colors~\cite{Moore:2014lga,Schmitz:2020syl}. 
In particular, the blue line corresponds to the parameters $(v_\phi^{},\lambda_{\phi S}^{})=(200~{\rm TeV},1.6)$ that produce the maximum amplitude. 
It is also noteworthy that there exists a small parameter region around $(v_\phi^{},\lambda_{\phi S}^{})=(1~{\rm TeV},1.1)$ that might be tested by SKA, as seen from the red line. 
%

\section{Dark Matter}\label{sec:DM}
In our model, $S$ is a natural DM candidate because it can not decay into SM particles due to the $Z_2^{}$ symmetry. 
Without a FOPT, the relic abundance of $S$ is determined by the usual freeze-out mechanism.   
However, the presence of a strong FOPT significantly changes this picture, as first discussed in Refs.~\cite{Baker:2019ndr,Chway:2019kft} in order to surpass the Griest–Kamionkowski bound~\cite{Griest:1989wd}.\footnote{
The heavy DM can be also produced by the bubble expansion with relativistic bubble wall velocity~\cite{Azatov:2021ifm}.  
See also Ref.~\cite{Azatov:2021irb} for applying the same mechanism to baryogenesis.   
}  
%
%
In the following, we represent the temperature-dependent vev (mass) of $\phi~(S)$ as $v_\phi^{}(T)$ ($m_S^{}(T)$) to distinguish it from that at zero temperature $v_\phi^{}$ ($m_S^{}(v_\phi^{})$).  
%

During the supercooling, $m_S^{}(T)$ is typically much larger than the kinetic energy $\sim T$.   
For example, when $T_n^{}/T_c^{}=10^{-3}$ and $\lambda_{\phi S}^{}\sim 1$, we have $m_S^{}(T_n^{})/T_n^{}\sim v_\phi^{}/T_n^{}\sim T_c^{}/T_n^{}=10^3$. 
In this case, $S$ particles can experience filtering-out effects such that some of them are reflected by bubble walls~\cite{Baker:2019ndr,Chway:2019kft}.  
The average number density of $S$ particles in the true vacuum is given by~\cite{Baker:2019ndr,Chway:2019kft,Hong:2020est}
\aln{
n_S^{\rm true}(T)&=\frac{1}{v_w^{}\gamma}\int \frac{d^3\mathbf{p}}{(2\pi)^3}\frac{-p_z^{}}{|\mathbf{p}|}\frac{1}{e^{\tilde{E}(\mathbf{p})/T}-1}\Theta(-p_z^{}-m_S^{}(T))\label{ntrue}
\\
&\sim -\frac{T^3}{4\pi^2}\left(\frac{\gamma (1-v_w^{})m_S^{}(T)/T+1}{\gamma^3(1-v_w^{})^2}\right)e^{-\gamma (1-v_w^{})\frac{m_S^{}(T)}{T}}~,\label{analytical ntrue}
}
where 
\aln{
\gamma=\frac{1}{\sqrt{1-v_w^2}}~,\quad \tilde{E}(\mathbf{p})=\gamma(|\mathbf{p}|+v_w^{}p_z^{})~,
}
and $\Theta(x)$ is the Heaviside step function. 
\begin{figure}[t!]
\begin{center}
\includegraphics[width=8cm]{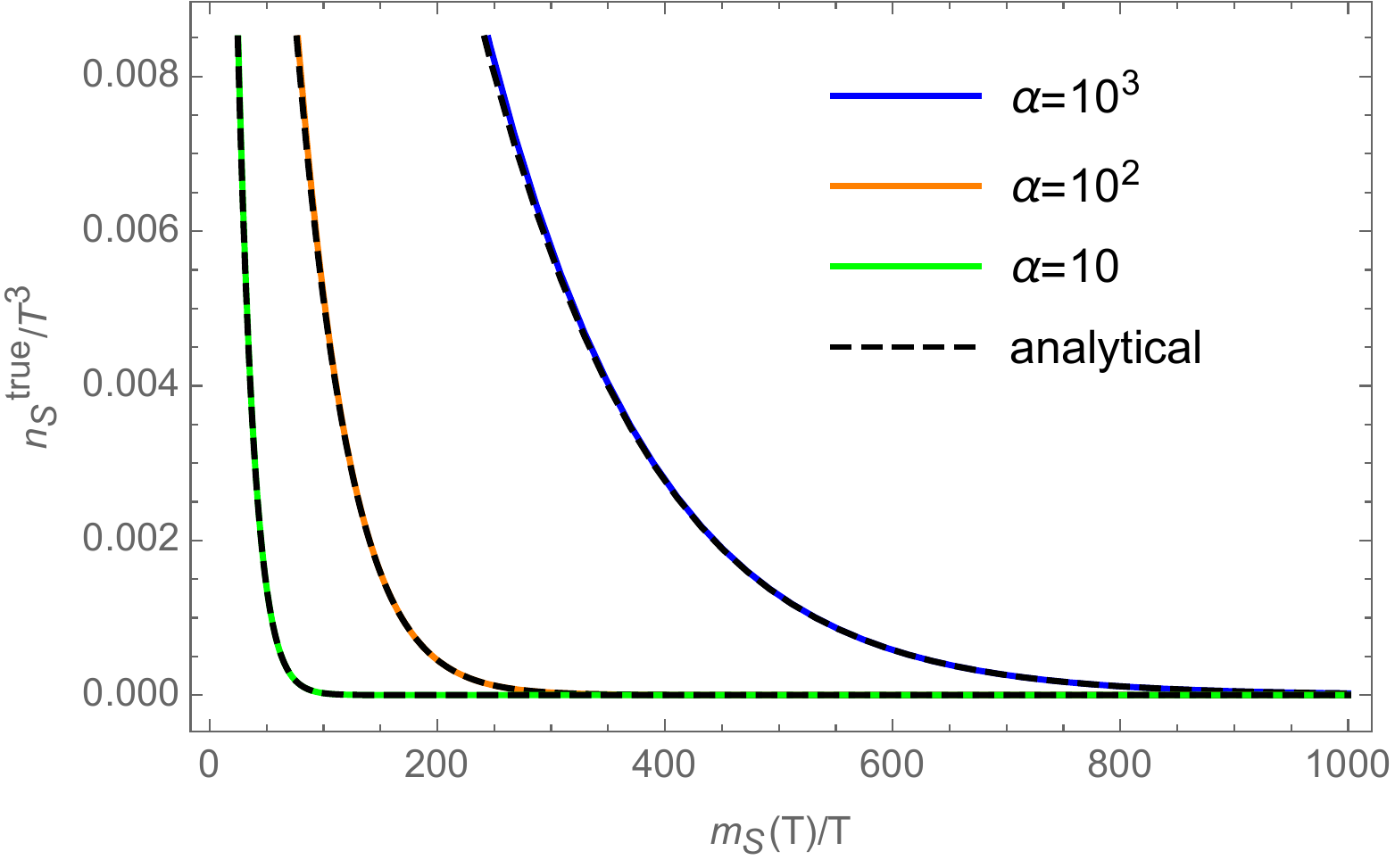}
\includegraphics[width=6cm]{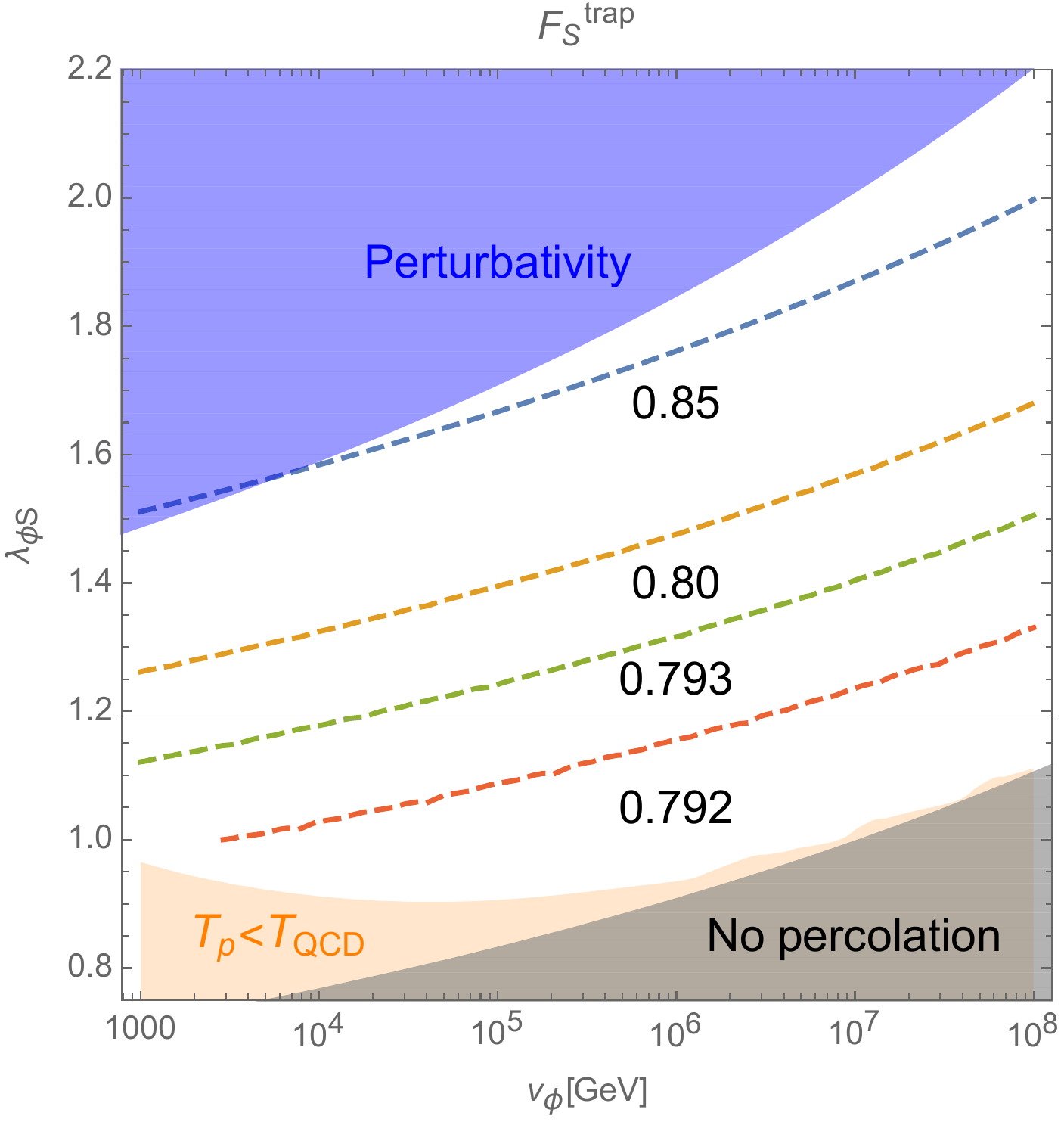}
\includegraphics[width=7cm]{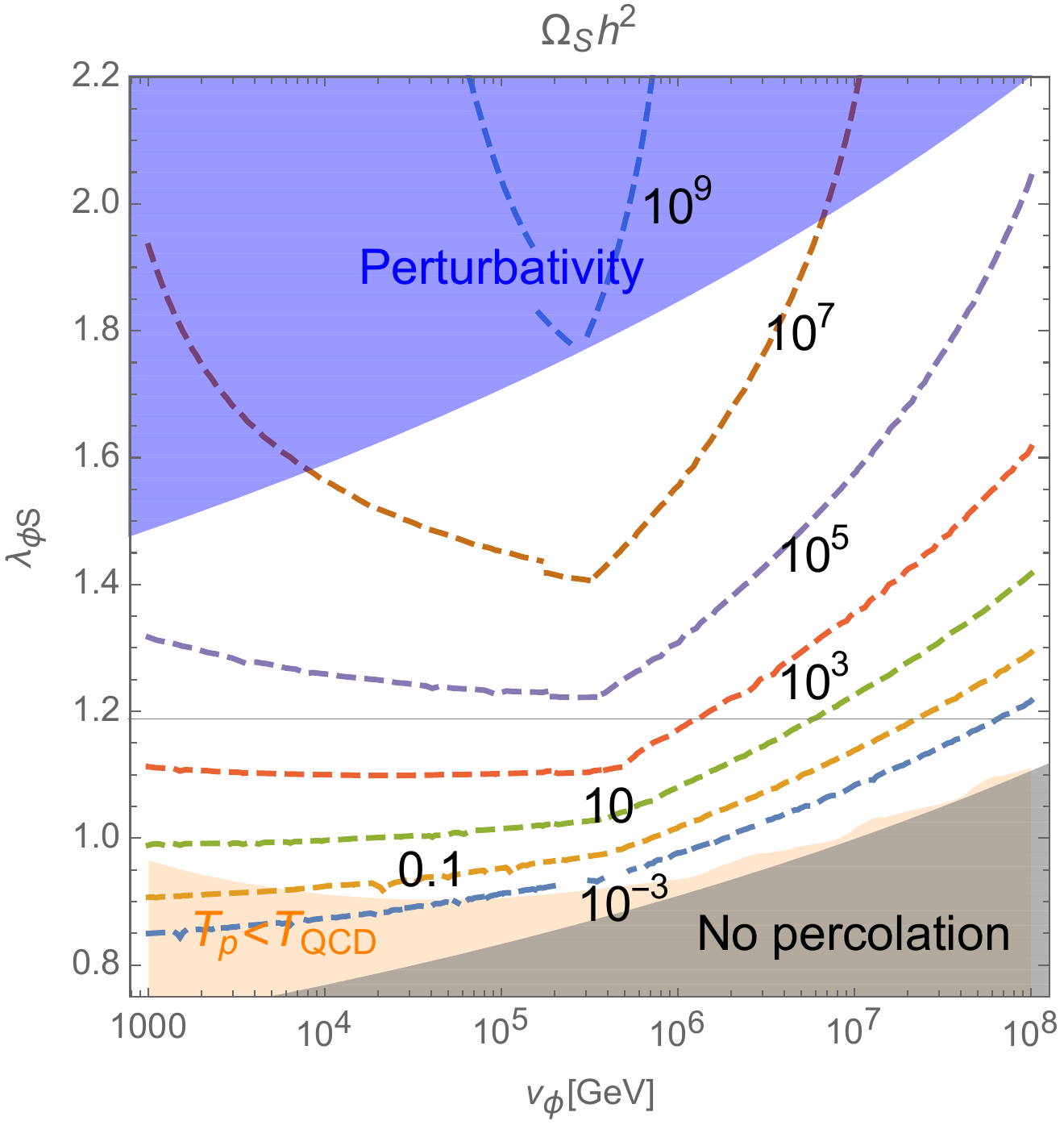}
\end{center}
\caption{Upper Left: Number density of $S$ in the true vacuum as a function of $m_S^{}(T)/T$.
Upper Right: Corresponding trapping fraction $F_S^{\rm trap}$. 
 Lower: Contours of $\Omega_S^{}h^2$ in the $v_\phi^{}$-$\lambda_{\phi S}^{}$ plane.
}
\label{fig:abundance}
\end{figure}
In the left panel in Fig.~\ref{fig:abundance}, we plot Eq.~(\ref{ntrue}) as a function of $m_S^{}(T)/T$ for different values of $v_w^{}$ in the case of the Jouguet detonation. 
The dashed black lines correspond to the analytical ones  Eq.~(\ref{analytical ntrue}).
%
%
%
%
Qualitatively, $n_S^{\rm true}$ is determined by the competition between $\gamma (1-v_w^{})$ and $m_S^{}(T)/T$ as we can see from the exponent in Eq.~(\ref{analytical ntrue}). 
%
%
It is also useful to define the trapping fraction as
\aln{F_S^{\rm trap}=1-\frac{n_S^{\rm true}}{n_S^{\rm false}}~,
} 
where $n_S^{\rm false}=\zeta(3)T^3/\pi^2$ is the number density of $S$ in the false vacuum.  
In the upper right panel in Fig.~\ref{fig:abundance}, we show the contours of $F_S^{\rm trap}$ in the $v_\phi^{}$-$\lambda_{\phi S}^{}$ plane. 
From this panel, one can see that $15\sim 20$~\% of $S$ particles penetrate the bubble walls.

Penetrated $S$ particles can no longer maintain thermal equilibrium because they are too heavy compared to the temperature, and they will survive until today. 
%
By taking the reheating era after the phase transition into account, the  relic abundance of $S$ at the present Universe is evaluated as
\aln{
\Omega_{S}^{}h^2&=\frac{m_S^{}(v_\phi^{})n_{S}^{\rm today}}{\rho_{c}^{}/h^2}\sim \frac{m_S^{}(v_\phi^{})}{\rho_c^{}/h^2}\frac{g_0^{}T_0^3}{g_R^{}T_R^{3}}\times \left(\frac{a(T_p^{})}{a(T_R^{})}\right)^3 \times n_S^{\rm true}(T_p^{})~,
\\
&\sim \frac{m_S^{}(v_\phi^{})}{3M_{\rm Pl}^2H_0^2/h^2}\frac{g_0^{}T_0^3}{g_R^{}T_R^{3}}\times \frac{\pi^2 g_R^{}T_R^4/30}{\epsilon}\times n_S^{\rm true}(T_p^{})~,\label{abundance of S}
}
where $T_0^{}=2.73~$K, $g_0^{}=3.9$, $\rho_c^{}=3M_{\rm Pl}^2H_0^2$, and $H_0^{}=100h~{\rm km}\cdot {\rm s}^{-1}\cdot {\rm Mpc}^{-1}$. 
When the reheating process is fast enough, $T_R^{}$ takes the maximum value
\aln{T_R^{}=T_\epsilon^{}=\left(\frac{30\epsilon}{\pi^2 g_R^{}}\right)^{1/4}=0.07\times \left(\frac{100}{g_R^{}}\right)^{1/4}m_S^{}(v_\phi^{})\ll m_S^{}(v_\phi^{})~,
}
which guarantees that the thermal productions of $S$ after the reheating are negligible. 
More detail calculations of $T_R^{}$ are presented in \ref{reheating}. 
%
In the lower panel in Fig.~\ref{fig:abundance}, we show the contours of $\Omega_S^{}h^2$ in the $v_\phi^{}$-$\lambda_{\phi S}^{}$ plane.  
Because the filtering effects are not much strong, the relic abundance of $S$ is typically overproduced, and only the small parameter region $\lambda_{\phi S}^{}\lesssim  1$ is allowed by  the current DM abundance.  
Note that in principle there is also a parameter region that is excluded by the XENON experiment~\cite{Aprile:2018dbl}, but it is not visible in Fig.~\ref{fig:abundance} because such a region exists in $v_\phi^{}<1$~TeV as long as $\lambda_{HS}^{}\lesssim 0.6$~\cite{Hamada:2021jls}.

\

We should comment on the fate of trapped particles. 
After $T=T_p^{}$, the false vacuum remnants continue to shrink due to the inward vacuum energy pressure, and the corresponding time scale is $\tau_{\rm shrink}^{}\sim H(T_p^{})^{-1}/v_w^{}\sim M_{\rm Pl}^{}/T_p^2$. 
On the other hand, the trapped scalar particles annihilate into SM particles via  $SS\rightarrow HH$ and $\phi\phi\rightarrow HH$, and the typical time scale of them is $\tau_{\rm anni}^{}\sim 16\pi\lambda_{\phi(S) H}^{-2}T_p^{-1}$, which is always much shorter than $\tau_{\rm shrink}^{}$ in the present model.
\footnote{In fact, when $v_\phi^{}=10^8~$GeV, $\lambda_{\phi H}^{}$ becomes $\sim 10^{-6}$ (see Eq.~(\ref{vev})), and $\tau_{\rm anni}^{}$ can become comparable to $\tau_{\rm shrink}^{}$ if a FOPT is moderate $T_p^{}\sim v_\phi^{}$. 
In our case, $T_p^{}$ is much smaller than $v_\phi^{}$, which implies $\tau_{\rm anni}^{}\ll \tau_{\rm shrink}^{}$. 
}  
Therefore, all the trapped scalar particles are rapidly  annihilating into the light (SM) particles, and no false vacuum remnants can survive.\footnote{These evaporation processes can also inject entropy to the plasma in the true vacuum, which can further reduce the relic abundance of $S$. Note also that similar trappings can happen for the SM particles with masses $\geq T_{\rm QCD}^{}$ when $T_p^{}\lesssim 100~{\rm GeV}$.  
But they rapidly annihilate or decay into light SM particles that can penetrate the bubble walls.      
}  
On one hand, this result is phenomenologically desirable because only the penetrated particles can contribute to the DM abundance. 
On the other hand, it is also a little disappointing because there are many interesting phenomena such as the formation of solitonic objects based on the above filtering mechanism~\cite{Krylov:2013qe,Hong:2020est,Kawana:2021tde,Gross:2021qgx,Asadi:2021yml}.   
We want to investigate these possibilities in future publications.  

\section{Conclusions} \label{sec:conclusion}
%
We have studied the cosmology in the minimal extended model that can realize the EW scale and DM  ~\cite{Haruna:2019zeu,Hamada:2020wjh,Hamada:2021jls}. 
In this model, the Universe can remain trapped in the false vacuum for a very long period of time due to the assumption of classical conformality.  
One of the observational signatures of such a strong FOPT is the stochastic GWs, and we have calculated the GW energy spectrum by taking the entropy productions via reheating into account.  
We found that the peak amplitude can become as large as $10^{-10}$ around $f\sim 10^{-3}~$Hz for some model parameters under the envelope approximation.   
Then, we have calculated the thermal relic abundance of additional scalar $S$.
%
Contrary to the usual freeze-out mechanism, $S$ particles can experience the filtering-out effects due to the huge mass gap between the false and true vacua, and only a small fraction of $S$ particles can penetrate the bubble walls. 
As a result, we found that only the small parameter region $0.8\lesssim \lambda_{\phi S}^{}\lesssim 1$ is allowed for the coexistence between FOPT and DM abundance. 
Although we have not investigated in this paper, there are many other interesting phenomena induced by a strong FOPT such as formation of primordial black holes or solitonic objects, non-thermal productions of DM, EW baryogenesis and so on. 
We would like to study these possibilities in the future.

\section*{Acknowledgements} 
We would like to thank Ke-Pan Xie, Philip Lu, Kengo Shimada, and Kei Yagyu for useful discussions and comments.    
This work is supported by Grant Korea NRF-2019R1C1C1010050, 2019R1A6A1A10073437. 
%

\appendix 
\def\thesection{Appendix \Alph{section}}

\section{Reheating temperature}\label{reheating}
We represent the decay (interaction) rate of $\phi$ as $\Gamma_\phi^{}$ in general.  
When $\Gamma_\phi^{}\gg H_{\epsilon}^{}$, the inflaton energy is immediately transferred to the decayed particles, and the reheating temperature is determined by 
\aln{\frac{\pi^2 g_R^{}}{30}T_R^4=\epsilon\quad \therefore\ T_R^{}=\left(\frac{30\epsilon^{}}{\pi^2g_R^{}}\right)^{1/4}=T_\epsilon^{} ~,
\label{TR 1}
}
where $g_\epsilon^{}$ is the effective number of dof at the reheating. 
On the other hand, when $\Gamma_\phi^{}\ll H_{\epsilon }^{}$, the reheating process is very slow and completed at round $t\sim \Gamma_\phi^{}$. 
%
Since the radiation energy becomes equivalent to that of $\phi$ at this moment, we have
\aln{
H(T_R^{})^2=\frac{2\rho_R^{}}{3M_{\rm Pl}^2}\sim 
\Gamma_\phi^{2}\quad \therefore\  T_R^{}\sim \left(\frac{45}{\pi^2 g_R^{}}\right)^{1/4}(M_{\rm Pl}^{}\Gamma_\phi^{})^{1/2}~. \label{TR 2}
}
%
%
See also Ref.~\cite{Garcia:2020wiy} for more detailed calculations of reheating temperature. Let us now consider our model.  
In the following, we represent the masses of scalar particles at zero temperature simply as $m_\phi^{},~m_S^{},~m_h^{}$.  
Because $m_\phi^{}\ll 2m_S^{}$, the main decay mode of $\phi$ is $H$, and the decay rate is qualitatively given by
\aln{\Gamma_\phi^{\rm decay}\sim \frac{(\lambda_{\phi H}^{}v_\phi^{})^2}{16\pi m_\phi^{}}=\frac{1}{16\pi}\frac{m_h^4}{m_\phi^{}v_\phi^2}~, 
}
where we have used Eq.~(\ref{vev}). 
From Eqs.~(\ref{VTI})(\ref{HTI}), the condition $\Gamma_\phi^{\rm decay}=H_{\epsilon}^{}$ is written as
\aln{M_{\rm Pl}^{}m_h^4=\frac{1}{4\pi}\left(\frac{2\lambda_{\phi S}^{}}{3}\right)^{1/2}m_S^{3}v_\phi^2~,
}
which determines the boundary value of $v_\phi^{}$ as 
\aln{v_\phi^{R}\sim (\lambda_{\phi S}^{-2}M_{\rm Pl}^{}m_h^4)^{1/5}\sim \lambda_{\phi S}^{-2/5}\times 2\times 10^5~{\rm GeV}. 
}  
Namely, when $v_\phi^{}\lesssim v_\phi^R$, $T_R^{}$ is given by Eq.~(\ref{TR 1}) while it becomes Eq.~(\ref{TR 2}) when $v_\phi^{}\gtrsim v_\phi^R$. 
By substituting $\Gamma_\phi^{\rm decay}$ into $\Gamma_\phi^{}$ in  Eq.~(\ref{TR 2}), we have
\aln{T_R^d=\frac{1}{4\pi}\left(\frac{20}{g_R^{}}\right)^{1/4}\left(\frac{M_{\rm Pl}^{}}{m_\phi^{}}\right)^{1/2}\frac{m_h^2}{v_\phi^{}}~.
}
Note that the decay $\phi\rightarrow HH$ is kinematically forbidden when $T>m_\phi^{}$ because the Higgs thermal mass $\sim y_t^{} T$ exceeds $m_\phi^{}$.  
To summarize, the reheating temperature determined by the decay is
\aln{T_R^{\rm decay}=\begin{cases}  T_R^I & \text{for }v_\phi^{}\lesssim v_\phi^R
\\
{\rm min}\{T_R^{d},~m_\phi^{}\}& \text{for }v_\phi^{}\gtrsim v_\phi^R
\end{cases}~.
}
In addition to the decay process, we also have to consider the scattering process $\phi H\rightarrow \phi H$, which is always kinematically allowed. 
The corresponding interaction rate is qualitatively given by
\aln{\Gamma_\phi^{\rm Scat}\sim \frac{\lambda_{\phi H}^2}{16\pi}\begin{cases}T   &  \text{for }T\gtrsim m_\phi^{}
\\
T^3/m_\phi^2 &  \text{for }T\lesssim m_\phi^{}
\end{cases}~.
}
By substituting this into $\Gamma_\phi^{}$ in  Eq.~(\ref{TR 2}) and solving it as a function of $T_R^{}$, we have
\aln{
T_R^{\rm Scat}\sim \begin{cases} \frac{\lambda_{\phi H}^2}{\pi^2}\sqrt{\frac{5}{32 g_R^{}}}M_{\rm Pl}:=T_R^{H} &  \text{for }T\gtrsim m_\phi^{}
\\
\frac{\pi^2}{\lambda_{\phi H}^{2}}\sqrt{\frac{32g_R^{}}{5}}\frac{m_\phi^2}{M_{\rm Pl}^{}}=m_\phi^{}\left(\frac{m_\phi^{}}{T_R^{H}}\right):=T_R^L &  \text{for }T\lesssim m_\phi^{}
\end{cases}~.
} 
We can see that only the $T\gtrsim m_\phi^{}$ case is physically meaningful because $T_R^L$ automatically exceeds $m_\phi^{}$ when $T_R^H<m_\phi^{}$, and this is inconsistent with the assumption $T\lesssim m_\phi^{}$.  
Moreover, $T_R^H$ has to satisfy 
\aln{T_R^{H}>m_\phi^{}\ \leftrightarrow\ M_{\rm Pl}^{}m_h^4>\pi\sqrt{\frac{2\lambda_{\phi S}^{}g_R^{}}{5}}v_\phi^4m_S^{}(v_\phi^{})~,
}
which is qualitatively the same condition as $v_\phi^{}\lesssim v_\phi^R$ as long as $\lambda_{\phi S}^{}={\cal O}(1)$. 
Therefore, the scattering effects are always irrelevant in our case, and the reheating temperature is determined by $T_R^{\rm decay}$.

\bibliography{Bibliography}
\bibliographystyle{utphys}

\end{document}